\RequirePackage{fix-cm}
\documentclass[twocolumn]{svjour3}          % twocolumn
\smartqed  % flush right qed marks, e.g. at end of proof
\usepackage{natbib}
\usepackage{multirow}
\usepackage{graphicx}
\usepackage{amsmath}
\usepackage{xcolor}
\colorlet{red}{white}

\newcommand{\mathcolorbox}[1]{#1}

\usepackage{ulem} % use normalem to protect \emph
\newcommand{\hl}[1]{#1}

\newcommand{\hz}[1]{#1}

\makeatletter
\newcommand\tabfill[1]{%
  \dimen@\linewidth
  \advance\dimen@\@totalleftmargin
  \advance\dimen@-\dimen\@curtab
  \parbox[t]\dimen@{#1\ifhmode\strut\fi}%
}
\makeatother

\makeatletter
\renewcommand*\env@matrix[1][\arraystretch]{%
  \edef\arraystretch{#1}%
  \hskip -\arraycolsep
  \let\@ifnextchar\new@ifnextchar
  \array{*\c@MaxMatrixCols c}}
\makeatother
\journalname{Rock Mechanics and Rock Engineering - Accepted Manuscript}
\begin{document}

\title{Transversely isotropic poroelastic behaviour of the Callovo-Oxfordian claystone: A set of stress-dependent parameters 
}

\titlerunning{Transversely isotropic poroelastic behaviour of the Callovo-Oxfordian claystone}        % if too long for running head

\author{Philipp Braun         \and
        Siavash Ghabezloo     \and
        Pierre Delage         \and
        Jean Sulem        \and
        Nathalie Conil       
}


\institute{           P. Braun         \and
        	S. Ghabezloo     \and
        	P. Delage         \and
        	J. Sulem 		\at
              Laboratoire Navier, 6-8 avenue Blaise-Pascal, Cité Descartes
77455 Champs-sur-Marne, Paris, France, philipp.braun@enpc.fr
           \and
              N. Conil        
			\at
              Andra, Meuse/Haute-Marne Underground Research Laboratory, Bure, France \\
  }

\date{Accepted: 5 October 2020}
\onecolumn

\maketitle

\begin{abstract}
In the framework of a deep geological radioactive waste disposal in France, the hydromechanical properties of the designated host rock, the Callovo-Oxfordian claystone (COx), are investigated in laboratory tests. Experiments presented in this study are carried out to determine several coefficients required within a transversely isotropic material model. They include isotropic compression tests, pore pressure tests, and deviatoric loading tests parallel and perpendicular to the bedding plane. We emphasize the adapted experimental devices and testing procedures, necessary to detect small strains under high pressures, on a material, which is sensitive to water and has a very low permeability. 
In particular, we discovered a significant decrease of elastic stiffness with decreasing effective stress, which was observed to be reversible. In both isotropic and deviatoric tests, a notable anisotropic strain response was found. The Young modulus parallel to bedding was about 1.8 times higher than the one perpendicular to the bedding plane. A notably low Poisson ratio perpendicular to the bedding plane with values between 0.1 and 0.2 was evidenced. While the anisotropy of the back-calculated Biot coefficient was found to be low, a significant anisotropy of the Skempton coefficient was computed. 
The performed experiments \hz{provide} an overdetermined set of material parameters at different stress levels. Using all determined parameters in a least square error regression scheme, seven independent elastic coefficients and their effective stress dependency \hz{are} characterized. 
\hz{Parameters measured under isotropic loading are well represented by this set of coefficients, while the poroelastic framework with isotropic stress dependency is not sufficient to describe laboratory findings from triaxial loading.}

\end{abstract}

\twocolumn

\section*{List of Symbols}
Note that the matrix notation is used throughout this work. 
\begin{tabbing}
\hspace{0.1\columnwidth}  \= \hspace{0.1\columnwidth}   \= \hspace{0.8\columnwidth}    \kill \ignorespaces
$\varepsilon_i$ \> \tabfill{Strain vector containing the \hz{six} independent components of the second rank strain tensor}\\
$M_{ij}$ \> Drained stiffness tensor in matrix format	\\
$C_{ij}$ \> Drained compliance tensor in matrix format	\\
$\sigma_i$ \> \tabfill{Stress vector containing the \hz{six} independent components of the second rank stress tensor}\\
$b_i$ \> Biot's coefficient for $i$-th direction	\\
$p_f$ \> Pore fluid pressure	\\
$E_{i}$ \> \tabfill{Drained Young's modulus in the $i$-th direction}			\\
$\nu_{zh}$ \> \tabfill{Drained Poisson's ratio perpendicular to bedding}			\\
$\nu'_{zh}$ \> \tabfill{Apparent drained Poisson's ratio under loading parallel to bedding}	\\
$\nu_{hh}$ \> Drained Poisson's ratio parallel to bedding			\\
$G$ \> Shear modulus perpendicular to bedding			\\
$G'$ \> Shear modulus parallel to bedding			\\
${\phi_0}$ \> Porosity		\\
$M$ \> Biot's undrained modulus $M$		\\
$N$ \> Biot's skeleton modulus $N$		\\
$K_{f}$ \> Bulk modulus of the pore fluid				\\
$K_{\phi}$ \> Unjacketed pore modulus				\\
$B_i$ \> Skempton's coefficient for $i$-th direction	\\
$K_{s}$ \> Unjacketed bulk modulus				\\
$M_{ij}^{u}$ \> Undrained stiffness tensor in matrix format	\\
$C_{ij}^{u}$ \> \tabfill{Undrained compliance tensor in matrix format}	\\
$E_{u,i}$ \> \tabfill{Undrained Young's modulus in the $i$-th direction}			\\
$\nu_{u,zh}$ \> \tabfill{Undrained Poisson's ratio perpendicular to bedding	}		\\
$\nu_{u,hh}$ \> \tabfill{Undrained Poisson's ratio parallel to bedding}			\\
$\sigma'$ \> Terzaghi isotropic effective stress\\
$\sigma$ \> Isotropic total stress	\\
$\varepsilon_v$ \> Volumetric strain \\
$K_{d}$ \> Isotropic drained bulk modulus			\\
$H$ \> Biot's pore pressure loading modulus			\\
$b$ \> Isotropic Biot's coefficient			\\
$V$ \> Specimen volume		\\
$m_f$ \> Pore fluid mass\\
$K_{u}$ \> Isotropic undrained bulk modulus			\\
$B$ \>  Isotropic Skempton's coefficient \\
$D_{i}$ \> \tabfill{Drained isotropic compression modulus in the $i$-th direction	}		\\
$U_{i}$ \> \tabfill{Undrained isotropic compression modulus in the $i$-th direction	}		\\
$H_{i}$ \> \tabfill{Biot's pore pressure loading modulus in the $i$-th direction}			\\
$R_{D}$ \> \tabfill{Anisotropy ratio in drained isotropic compression}		\\
$R_{U}$ \> \tabfill{Anisotropy ratio in undrained isotropic compression }				\\
$R_{H}$ \> \tabfill{Anisotropy ratio in pore pressure loading 	}	\\
$R_{E}$ \> \tabfill{Anisotropy ratio of drained Young's moduli} 		\\
$q$ \> Deviatoric stress			\\
$E_z^{\infty}$, $E_z^{0}$, $\beta$ \>  \>  \tabfill{Model parameters for regression analysis}		\\
$\rho$ \> Wet density\\
$\rho_d$ \> Dry density\\
$w$ \> Water content\\
$S_r$ \> Saturation degree\\
$s$ \> Suction\\
$S$ \> Standard deviation of the estimate\\
\end{tabbing}

\section{Introduction}
\label{sec:el:1}
The French National Agency for Nuclear Waste Management, Andra, is investigating the possibility of constructing a deep geological nuclear waste disposal facility in \hl{the} Callovo-Oxfordian (COx) claystone, located in the east of France. 

To study the \hl{host} material \hl{and demonstrate the feasibility of the project}, Andra constructed an underground research laboratory (URL) at 490 m depth, \hl{where} the claystone has an average porosity of 17.5 \% and an average water content of 7.9 \% \citep{Conil201861}. An average clay content around 42 \% was \hl{mentioned} by the same authors. \cite{Wileveau200786} determined a vertical and a minor horizontal total stress close to 12 MPa, a major horizontal stress close to 16 MPa and a pore pressure of 4.9 MPa. 

During extensive laboratory studies on its hydro-mechanical behaviour, the COx claystone was found transversely isotropic with larger stiffness parallel to the bedding plane \citep{Chiarelli2000,Escoffier2002,Andra2005,Mohajerani201211,Zhang201279,Belmokhtar201787}. This feature, \hl{revealed} also in many other types of shales, can be attributed to the sedimentary history of the material. \hl{Due to this} fact, transversely isotropic poroelasticity \hl{was chosen as a framework} for hydromechanical modelling purposes. A larger number of experiments is therefore necessary\hl{, compared} to \hl{an isotropic material, to} capture the complete material behaviour (e.g. two triaxial tests have to be carried out to determine the Young moduli and Poisson ratios both perpendicular and parallel to the bedding plane).
While there are extensive data on the Young moduli \hl{perpendicular to bedding,} provided by the aforementioned authors, \hl{fewer} data \hl{are} given \hl{on the Young modulus parallel to bedding and} the  Poisson ratios of saturated COx \citep{Menaceur201529,Belmokhtar201819}. Several experiments for establishing a set of poroelastic parameters were carried out in the present work. This includes triaxial compression parallel and perpendicular to the bedding plane in drained and undrained conditions, isotropic compression tests in both drained and undrained conditions, and pore pressure loading tests under constant total stress. \hl{The} anisotropic strain response \hl{was measured in all experiments.}

The COx claystone can be considered as a swelling claystone \hl{\mbox{\citep{Schmitt199441}}. In its natural state, the claystone is fully saturated. Therefore, swelling can only be induced when the material is brought in contact with water after desaturation, or when it is submitted to an effective stress release.} Several authors working on this rock reported an increase of stiffness and mechanical resistance with decreasing water content. Even though \hl{specimens} are well protected from drying during all handling phases from core extraction to laboratory testing, they desaturate slightly, with a saturation degree between 90 and 95 \% generally observed. 
To evaluate parameters on \hl{specimens} under well defined saturated conditions, the specimens have to be hydrated in the testing apparatus. Special attention was paid here, as the application of stress during this phase is indispensable for avoiding free swelling and limiting \hl{specimen} damage. Due to the low permeability of the claystone, the saturation time might become very long, which is why we used adapted testing devices, allowing \hl{us} to reduce the drainage length of the specimens \hl{to 10 mm} \citep{Tang200845,Belmokhtar201787,Belmokhtar201819}. \hl{Improved testing procedures, presented in detail by \mbox{\cite{Braun2019}} were followed, comprised of step loading, which enable a better distinction between drained and undrained specimen state in the experiments. These time efficient procedures permit us to evaluate several poroleastic coefficients in each loading step, and hence to analyse inter-compatible parameters as a function of the isotropic effective stress.}

\hl{We refer here also to our companion paper \mbox{\citep{Braun2020}}, in which we investigated the thermal properties of the COx claystone under constant stress conditions. The present work was carried out under isothermal conditions, without any influence of thermal effects. However, in the companion paper we discuss thermal characteristics in undrained conditions, which are strongly coupled with the hydromechanical properties described in the following. }

\section{Poro-Elastic Framework}
\label{sec:el:2}

In this work we analyse the mechanical response of a porous rock due to external stress and pore pressure loading, based on the poroelastic constitutive equations, first described by \cite{Biot195759}. For a \hl{transversely} isotropic material one finds identical parameters for both directions $x=y=h$ parallel to the bedding plane and different properties in the direction $z$ (perpendicular to the bedding plane). The \hl{transversely} isotropic poroelastic stress-strain relationship can be written as \citep{Cheng199719}:

\begin{equation}
{\rm{d}}{\sigma _i} = {M_{ij}}{\rm{d}}{\varepsilon _j} + {b_i}{\rm{d}}p_{f}
\label{eq:el:el}
\end{equation}
where \hl{${\sigma _i}$ and $\varepsilon_i$ denote the stress and strain tensors, respectively, $M_{ij}$ the stiffness matrix, $b_i$ the Biot coefficients in \hz{the} $i$ direction and $p_{f}$ the pore fluid pressure. The strain tensor ${\sigma _i}$ is composed of:}
\begin{equation}
\sigma_i = {\left[ {{\sigma_x},{\sigma_y},{\sigma_z},{\sigma_{xy}},{\sigma_{yz}},{\sigma_{zx}}} \right]^\top}
\end{equation}

\begin{figure}[htbp]
  \includegraphics[width=\columnwidth]{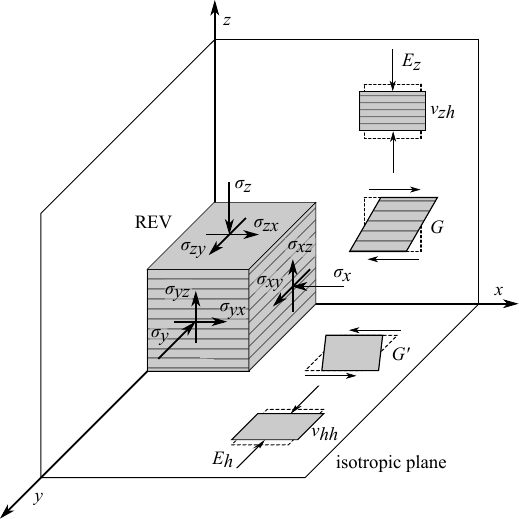}
\caption{\hl{Representative elementary volume in a transversely isotropic frame, adopted from \mbox{\cite{Popov201920}}. The $x-y$ plane corresponds to the isotropic plane. The $z$ axis is oriented perpendicular\hz{ly} to the isotropic plane. Different material properties $E_i$, $\nu_i$, $G$ and $G'$ describe the material response within the material planes, where the parameters in the $y-z$ plane (not displayed here) are equal to the ones in the $z-x$ plane.}}
\label{fig:el:trans_isotropy}       % Give a unique label
\end{figure}

\hl{The planes and directions in which the different stress tensor components and the material properties intervene is schematically represented in Fig. \ref{fig:el:trans_isotropy}.}
The strains in different directions are contained within the strain vector $\varepsilon_i$:
\begin{equation}
{\varepsilon}_i = {\left[ {{\varepsilon_x},{\varepsilon_y},{\varepsilon_z},{\varepsilon_{xy}},{\varepsilon_{yz}},{\varepsilon_{zx}}} \right]^\top}
\end{equation}

The vector $b_i$ represents the Biot effective stress coefficents in both directions of anisotropy:
\begin{equation}
b_i = {\left[ {{b_h},{b_h},{b_z},0,0,0} \right]^\top}
\label{eq:el:bi}
\end{equation}

We can also write the stiffness matrix $M_{ij}$ as follows:
\begin{equation}
\mathcolorbox{
M_{ij} =  
\begingroup % keep the change local
\setlength\arraycolsep{5pt}
\begin{pmatrix}[1.5]
{{M_{11}}}&{{M_{12}}}&{{M_{13}}}&0&0&0\\
{{M_{12}}}&{{M_{11}}}&{{M_{13}}}&0&0&0\\
{{M_{13}}}&{{M_{13}}}&{{M_{33}}}&0&0&0\\
0&0&0&2G'&0&0\\
0&0&0&0&{2G}&0\\
0&0&0&0&0&{2G}
\end{pmatrix}
\endgroup
}
\label{eq:el:M_matrix} 
\end{equation}
\begin{equation}
M_{11} = \frac{{{E_h}\left( {{E_z} - {E_h}{\nu_{zh}}^2} \right)}}{{\left( {1 + {\nu_{hh}}} \right)\left( {{E_z} - {E_z}{\nu_{hh}} - 2{E_h}{\nu_{zh}}^2} \right)}}
\label{eq:el:M11} 
\end{equation}
\begin{equation}
M_{12} = \frac{{{E_h}\left( {{E_z}{\nu_{hh}} + {E_h}{\nu_{zh}}^2} \right)}}{{\left( {1 + {\nu_{hh}}} \right)\left( {{E_z} - {E_z}{\nu_{hh}} - 2{E_h}{\nu_{zh}}^2} \right)}}
\end{equation}
\begin{equation}
M_{13} = \frac{{{E_h}{E_z}{\nu_{zh}}}}{{{E_z} - {E_z}{\nu_{hh}} - 2{E_h}{\nu_{zh}}^2}}
\end{equation}
\begin{equation}
M_{33} = \frac{{{E_z}^2\left( {1 - {\nu_{hh}}} \right)}}{{{E_z} - {E_z}{\nu_{hh}} - 2{E_h}{\nu_{zh}}^2}}
\label{eq:el:M33} 
\end{equation}
\begin{equation}
\mathcolorbox{
G' = \frac{M_{11}-M_{12}}{2}
}
\label{eq:el:G_shear} 
\end{equation}
where ${E_z}$ and $\nu_{zh}$ are the Young modulus and the Poisson ratio perpendicular to the bedding plane, and ${E_h}$ and $\nu_{hh}$ parallel to the bedding plane, respectively. \hl{$G$ denotes the shear modulus perpendicular to bedding and $G'$ the shear modulus parallel to bedding.} 

\hl{By using the inverse of the stiffness matrix $M_{ij}^{-1}=C_{ij}$, denoted as the drained compliance matrix, one can write:}
\begin{equation}
\label{eq:el:1}
{\rm{d}}{\varepsilon _j}{\rm{ = }}{C_{ij}}{\rm{d}}{\sigma _i} - {C_{ij}}{b_i}{\rm{d}}p_f
\end{equation}

\hl{The compliance matrix $C_{ij}$ is written as:}
\begin{equation}
\mathcolorbox{
C_{ij} =  
\begingroup % keep the change local
\setlength\arraycolsep{5pt}
\begin{pmatrix}
{{C_{11}}}&{{C_{12}}}&{{C_{13}}}&0&0&0\\
{{C_{12}}}&{{C_{11}}}&{{C_{13}}}&0&0&0\\
{{C_{13}}}&{{C_{13}}}&{{C_{33}}}&0&0&0\\
0&0&0&{1/(2G')}&0&0\\
0&0&0&0&{1/(2G)}&0\\
0&0&0&0&0&{1/(2G)}
\end{pmatrix}
\endgroup
}
\end{equation}
with 
\begin{equation}
\label{eq:entries_C}
\mathcolorbox{
\begin{array}{ll}
C_{11} &= 1/E_h	\\
C_{12} &= -\nu_{hh}/E_h	\\
C_{13} &= -\nu_{zh}/E_z	\\
C_{33} &= 1/E_z	\\
G'&=E_h/(1+\nu_{hh})	\\
\end{array}
}
\end{equation}

\hl{In laboratory experiments, the Young moduli and Poisson ratios are generally measured in a triaxial cell under deviatoric loads. The properties $E_{z}$ and $\nu_{zh}$ are determined on a specimen which is submitted to a deviatoric load perpendicular to the bedding plane $\sigma_z$, generating strains $\varepsilon_z$ and $\varepsilon_x=\varepsilon_y$ (Fig. \ref{fig:el:straingages}). We can then measure $E_z=\mathrm{d} \sigma_z / \mathrm{d} \varepsilon_z$ and $\nu_{zh}=-\mathrm{d} \varepsilon_x / \mathrm{d} \varepsilon_z$. A specimen\hz{,} which is loaded in \hz{the direction} parallel to the bedding plane (e.g. in direction of $y$) is required for the remaining properties. Here one is able to observe an axial strain $\varepsilon_y$ and two different radial strains $\varepsilon_x$ and $\varepsilon_z$. This provides $E_h=\mathrm{d} \sigma_y / \mathrm{d} \varepsilon_y$, $\nu_{hh}=-\mathrm{d} \varepsilon_x / \mathrm{d} \varepsilon_y$. The measured parameter $\nu_{zh}'=-\mathrm{d} \varepsilon_z / \mathrm{d}\varepsilon_y$ allows us to evaluate of $\nu_{zh}=\nu_{zh}'E_z/E_h$.
To be able to determine the shear modulus $G$ perpendicular to the isotropic plane, a third triaxial test is required, where a deviatoric load inclined with respect to the bedding plane is applied. Such loading would result in shear stresses perpendicular to the bedding plane, mobilizing $G$, but also inducing inhomogeneous stress and strain distributions within a specimen, which have to be considered in the analysis. 
While the shear modulus $G'$ is related to the other elastic coefficients (Eq. (\ref{eq:el:G_shear})), $G$ is an independent parameter. The evaluation of $G$ was not carried out in this work and is not further discussed here.}

\hl{The porosity variation $\mathrm{d}\phi $ is given by:}
\begin{equation}
\mathcolorbox{
{\rm{d}}\phi  =  - {b_i}{\rm{d}}{\varepsilon _i} + \frac{1}{N}{\rm{d}}p_f 	,\quad	 \frac{1}{N}=\frac{1}{M}-\frac{\phi_0}{K_f}
}
\label{eq:el:phi}
\end{equation}
\hl{where $M$ is Biot's undrained modulus, $N$ Biot's skeleton modulus and ${K_f}$ the pore fluid bulk modulus. Note that the expression of porosity variation needs an additional poroelastic parameter with respect to the ones presented previously for the stress-strain relation.
The Biot modulus $M$ can be expressed according to \hz{\mbox{\cite{Aichi201233}}} by:}
\begin{equation}
\mathcolorbox{
\begin{split}
\frac{1}{M} = 2(1 - {b_h})\left[ {\frac{{\left( {1 - {\nu _{hh}}} \right){b_h}}}{{{E_h}}} - \frac{{{\nu _{zh}}{b_z}}}{{{E_z}}}} \right] + \\
+ \frac{{\left( {1 - {b_z}} \right)}}{{{E_z}}}\left( {{b_z} - 2{\nu _{zh}}{b_h}} \right) + \phi_0 \left( {\frac{1}{{{K_f}}} - \frac{1}{{{K_\phi }}}} \right)
\end{split}
}
\label{eq:el:M}
\end{equation}
with ${K_\phi}$ representing the \hl{unjacketed pore modulus \mbox{\citep{Brown1975}}.}
In undrained conditions, the fluid mass \hl{remains constant and we can write ${\rm{d}} m_f= {\rm{d}} \left( \rho_f \phi_0 \right) =0$. Together with ${\rm{d}} \rho_f / \rho_f= {\rm{d}} p_f /K_f$ and Eq. (\ref{eq:el:phi}), we obtain an expression for the pore pressure change in undrained conditions:}
\begin{equation}
{\rm{d}}{p_f} ={M} {b_i}{\rm{d}}{\varepsilon _i}
\label{eq:el:dp_undrained}
\end{equation}

\hl{Using Eq. (\ref{eq:el:el}), one can express the pore pressure change ${\rm{d}} p_f$ (Eq. (\ref{eq:el:dp_undrained})) as a function of total stress change:}
\begin{equation}
{\rm{d}}{p_f} =\frac{1}{3} {B_i}{\rm{d}}{\sigma _i}
\label{eq:el:Bi_loading}
\end{equation} 
\hl{where: }
\begin{equation}
\mathcolorbox{
{B_i} = \frac{3 b_{j} C_{ij}}{\frac{1}{M}+b_i C_{ij} b_j } 	
}	
\label{eq:el:Bi_}
\end{equation} 

\hl{${B_i}$ comprises the anisotropic components of the Skempton coefficient:}
\begin{equation}
\mathcolorbox{
B_i = {\left[ {{B_h},{B_h},{B_z},0,0,0} \right]^\top} 
}	 
\end{equation} 

\hl{$B_z$ is the Skempton coefficient parallel to bedding and $B_h$ the Skempton coefficient perpendicular to bedding, arising from the mathematical framework of transverse isotropy. The classical parameter $B$ used for isotropic materials can here only describe the change of pore pressure under isotropic loading, for which $B=\sum B_i/3$. Under deviatoric loading however, the loading direction influences the amount of generated pore pressure, due to the anisotropic characteristics. Imagine a material with $B_z$ higher than $B_h$. According to Eq. (\ref{eq:el:Bi_loading}), if  this material is loaded in the $z$ direction, it generates a higher pore pressure increase than when loaded \hz{by} the same amount in the $h$ direction.}
%According to \cite{Cheng199719}, $B_i$ can be calculated as:
%\begin{equation}
\mathcolorbox{
%{B_i} = \frac{{3\sum\limits_j {{C_{ij}}}  - \frac{1}{K_s}}}{\sum\limits_{i} \sum\limits_{j} {{C_{ij}} - \frac{1}{{{K_s}}} + \phi_0 \left( {\frac{1}{{{K_f}}} - \frac{1}{{{K_s}}}} \right)}}
}
%\label{eq:el:Bi_general}
%\end{equation} 
%where the parameter $K_s$ represents the unjacketed bulk modulus \citep{Gassmann1951,Brown1975}. %Note that $K_s$ describes the isotropic deformation of the solid phase due to the assumption of micro-isotropy.

Analogously to the drained stiffness matrix, one can determine the undrained stiffness matrix \hl{$M^u_{ij}$ with \mbox{\citep{Cheng199719}}:}
\begin{equation}
M_{ij}^u = {M_{ij}} + M{b_i}{b_j}
\label{eq:el:Mu}
\end{equation} 
The inverse of the undrained stiffness matrix is the undrained compliance matrix $C^u_{ij}$, which provides the undrained Young's moduli $E_{u,i}$ and the undrained Poisson's ratios $\nu_{u,i}$:
\begin{equation}
\begin{array}{l}
{C_{11}^u} = 1/{E_{u,h}}\\
{C_{12}^u} =  - {\nu _{u,hh}}/{E_{u,h}}\\
{C_{13}^u} =  - {\nu _{u,zh}}/{E_{u,z}}\\
{C_{33}^u} = 1/{E_{u,z}}
\end{array}
\label{eq:el:Eu_nuu}
\end{equation} 

For this transversely isotropic relationship under micro-heterogeneity \hl{and micro-isotropy assumptions, \hz{seven}} material coefficients (i.e. $E_h$, $E_z$, $\nu_{hh}$, $\nu_{zh}$, $b_h$, $b_z$, $G$) \hl{describe the stress strain relationship, while \hz{three} additional parameters ($K_{\phi}$, $\phi_0$, $K_{f}$) are required for the porosity change \mbox{\citep{Aichi201233}}.} 

\subsection{Isotropic stress conditions}
\label{sec:el:poroelastic_iso}

\hl{When analysing experiments under isotropic stress conditions, it can be helpful to utilize some poromechanical relationships between volume changes and transversely isotropic deformations. 
Note that due to the anisotropic framework, the bulk parameters discussed in this section are \hz{unable} to describe a volume change under mean stress changes in general. 
Contrary to an isotropic material framework, the parameters can here only be related to isotropic stress changes.
Nevertheless, data from isotropic tests can provide complementary information to data from deviatoric stress tests, useful for inferring the complete set of elastic properties.}

Under isothermal conditions and isotropic stresses, the volumetric strain $\varepsilon_v$ is described with respect to the changes in isotropic Terzaghi effective stress $\sigma' = \sigma - p_f$ (where $\sigma$ is the total confining stress and $p_f$ the pore fluid pressure) as a sum of partial derivatives:
\begin{equation}
\mathcolorbox{
{\rm{d}}{\varepsilon _v} = \sum {\rm{d}}{\varepsilon _i} =\frac{1}{{{K_d}}}{\rm{d}}{\sigma'} + \frac{1}{{{K_s}}}{\rm{d}}{p_f} 
}
\label{eq:el:1}
\end{equation}
in which $K_d$ is the \hl{isotropic} drained bulk modulus \hl{and $K_s$ the unjacketed bulk modulus \mbox{\citep{Gassmann1951,Brown1975}}.} We are able to measure the Biot modulus $H$ through a change of pore pressure under constant confining stress, defined as:
\begin{equation}
\frac{1}{H} = \frac{1}{{{V_0}}}{\left( {\frac{{\partial V}}{{\partial {p_f}}}} \right)_{\sigma}} = \frac{1}{{{K_d}}} - \frac{1}{{{K_s}}}
\label{eq:el:2}
\end{equation}

The \hl{isotropic} Biot coefficient $b$ is defined as:
\begin{equation}
b = 1- \frac{K_d}{K_s}=\frac{K_d}{H}
\label{eq:el:b}
\end{equation}

In undrained conditions, the mass of pore fluid remains constant, which results in a change in volumetric strain and pore pressure through isotropic stress, as follows: 
\begin{equation}
{\rm{d}}{\varepsilon _v} = \frac{1}{{{K_u}}}{\rm{d}}\sigma  ,\quad
{\rm{d}}{p_f} = B{\rm{d}}\sigma  
\label{eq:el:3}
\end{equation}
Hereby $K_u$ is the \hl{isotropic} undrained bulk modulus and $B$ the \hl{isotropic} Skempton coefficient \hl{$B=\sum B_i/3$} \citep{Skempton195414}. 
Following the approach of \cite{Belmokhtar201787}, the moduli $D_i$, $U_i$ and $H_i$, which describe the anisotropic strain response to isotropic loads, are defined for a drained isotropic compression:
\begin{equation}
D_i = \left( \frac{{\rm{d}}{\sigma}}{{\rm{d}}{\varepsilon _i}} \right)_{{\rm{d}}p_f=0}
\label{eq:el:Ddef}
\end{equation}
for an undrained isotropic compression:
\begin{equation}
U_i = \left( \frac{{\rm{d}}{\sigma}}{{\rm{d}}{\varepsilon _i}} \right)_{{\rm{d}}m_f=0}
\label{eq:el:Udef}
\end{equation}
and for a pore pressure loading:
\begin{equation}
\mathcolorbox{
H_i = - \left( \frac{{\rm{d}}{p_f}}{{\rm{d}}{\varepsilon _i}} \right)_{{\rm{d}}\sigma=0}
}
\label{eq:el:Hdef}
\end{equation}

We can also define the anisotropy ratios between deformations in $h$ and in $z$ direction with $R_D=D_h/D_z$, $R_U=U_h/U_z$ and $R_H=H_h/H_z$, which provides:
\begin{equation}
D_h = K_d \left(2+ {R_D} \right)	,\quad	D_z = K_d \left(1+ \frac{2}{R_D} \right)  
\label{eq:el:Di}
\end{equation}
\begin{equation}
H_h = H \left(2+ {R_H} \right)	,\quad	H_z = H \left(1+ \frac{2}{R_H} \right)
\label{eq:el:Hi}
\end{equation}

The undrained moduli $U_i$ are linked to the drained ones with the relationship:
\begin{equation}
\frac{1}{U_i} = \frac{1}{D_i} -B \frac{1}{H_i} 
\label{eq:el:UiDi}
\end{equation}

One is able to deduce a relationship between the parameters from isotropic and deviatoric experiments: 
\begin{equation}
E_z = D_z \left(1- 2 \nu_{zh} \right)	,\quad	E_h = \frac {1-\nu_{hh}}{\frac{1}{D_h}+\frac{\nu_{zh}}{E_z}}
\label{eq:el:Ei}
\end{equation}

Also the parallel and perpendicular Biot's coefficients $b_{h}$ and $b_{z}$ can be calculated concurrently \citep{Belmokhtar201787}:
\begin{equation}
b_h = \frac{
{{\nu_{zh}} \mathord{\left/{\vphantom {1 {{H_z}}}} \right. \kern-\nulldelimiterspace} {{H_z}}}+
{1 \mathord{\left/{\vphantom {1 {{H_h}}}} \right. \kern-\nulldelimiterspace} {{H_h}}}}{
{{\nu_{zh}} \mathord{\left/{\vphantom {1 {{D_z}}}} \right. \kern-\nulldelimiterspace} {{D_z}}}+
{1 \mathord{\left/{\vphantom {1 {{D_h}}}} \right. \kern-\nulldelimiterspace} {{D_h}}}}
\label{eq:el:bh}
\end{equation}
\begin{equation}
b_z = 2 \nu_{zh} b_h + \left( 1-2\nu_{zh}   \right)  \frac{ D_z }{H_z}
\label{eq:el:bz}
\end{equation}
The \hl{bulk} Skempton's coefficient $B$ can be determined as:
\begin{equation}
B = \frac{{{K_u} - {K_d}}}{{b{K_u}}}
\label{eq:el:B}
\end{equation}
and the expression (\ref{eq:el:Bi_}) for the two Skempton coefficients $B_z$ and $B_h$ is simplified to:
\begin{equation}
{B_i} = \frac{{3\sum\limits_j {{C_{ij}}}  - \frac{1}{{{K_s}}}}}{{ {\frac{1}{{{K_d}}} - \frac{1}{{{K_s}}}}}}B
\label{eq:el:Bi}
\end{equation} 

\hl{Note that $K_{\phi}$ has to be determined separately in an unjacketed test by measuring the change in pore fluid mass during equal increments of pore pressure and total stress change (e.g. \mbox{\citealp{Gassmann1951}}; \mbox{\citealp{Brown1975}}; \mbox{\citealp{Coussy2004}}; \mbox{\citealp{Ghabezloo200814}}; \mbox{\citealp{Makhnenko2017}}). This fluid mass change is generally very small for common geomaterials, making the precise measurement a challenging task. For homogeneous materials at the micro-scale, one can assume $K_{\phi}=K_{s}$ \mbox{\citep{Berryman199217}}. If this is not known a priori, it is practical to use some poromechanical relationships for calculating $K_{\phi}$, as given by \mbox{\cite{Ghabezloo200814}}, which provide: }

\begin{equation}
\frac{1}{{{K_\phi }}} = \frac{1}{{{K_f}}} - \frac{{\left( {{1 \mathord{\left/
 {\vphantom {1 {{K_u}}}} \right.
 \kern-\nulldelimiterspace} {{K_u}}} - {1 \mathord{\left/
 {\vphantom {1 {{K_d}}}} \right.
 \kern-\nulldelimiterspace} {{K_d}}} + {1 \mathord{\left/
 {\vphantom {1 H}} \right.
 \kern-\nulldelimiterspace} H}} \right)}}{{\phi_0 H\left( {{1 \mathord{\left/
 {\vphantom {1 {{K_d}}}} \right.
 \kern-\nulldelimiterspace} {{K_d}}} - {1 \mathord{\left/
 {\vphantom {1 {{K_u}}}} \right.
 \kern-\nulldelimiterspace} {{K_u}}}} \right)}}
\label{eq:el:Kphi}
\end{equation}

The \hl{bulk modulus of the pore space} $K_{\phi}$ can hence be \hl{calculated by an experimental evaluation of the undrained bulk modulus} $K_{u}$, which is less tedious to be determined in an undrained isotropic compression test. The porosity $\phi_0$ can be obtained by volume and mass measurements of oven-dried specimens, whereas the bulk modulus of the pore fluid $K_{f}$ is generally known for a specific fluid, such as for example water. 
If one is able to measure more than the minimum \hl{number of \hz{eight} independent coefficients (i.e. $E_h$, $E_z$, $\nu_{hh}$, $\nu_{zh}$, $b_h$, $b_z$, $G$, $K_{\phi}$, given that $\phi_0$ and $K_{f}$ are known) experimentally,} an overdetermined set of parameters is obtained, which can help to check for a compatibility of these values. \hl{In this study we did not investigate $G$, therefore the parameter set which we aimed to characterize contains \hz{seven} independent coefficients.}

\section{Materials and Methods}
\label{sec:el:3}
\subsection{Specimen preparation}
\label{sec:el:3.1}
In this laboratory study we investigated specimens trimmed from COx cores that were extracted at the URL from horizontal boreholes.
To preserve the material in its in-situ state as best as possible, that means to avoid desaturation \citep{Ewy201538} and mechanical damage, cores were shipped and stored in T1 cells, developed by Andra \citep{Conil201861}. \hl{Each} cell protects an 80 mm diameter COx core, which is wrapped in aluminium foil and a latex membrane to prevent drying. A plastic tube is placed around the core and cement is cast between the membrane and the tube, forming a layer of mechanical protection. This minimises, together with a spring system in axial direction, volume changes of the core. 

\begin{figure*}[htbp]
  \includegraphics[width=0.7\textwidth]{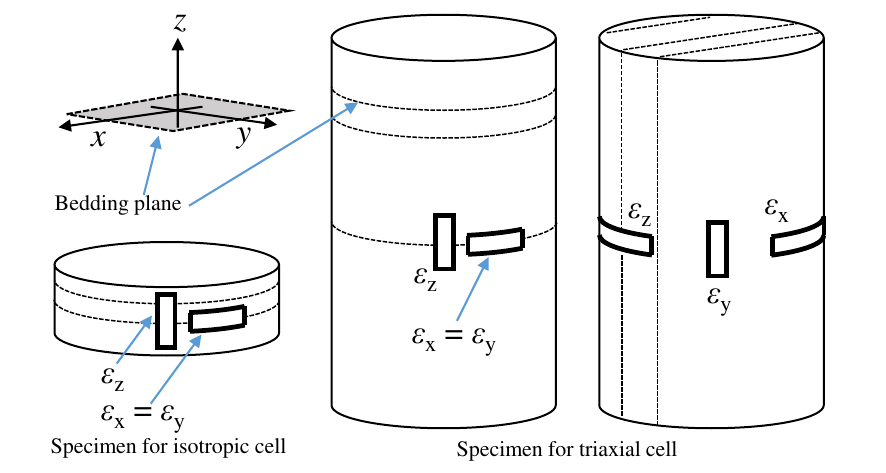}
\caption{Strain gages, oriented and attached with respect to the bedding plane of the COx specimens.}
\label{fig:el:straingages}       % Give a unique label
\end{figure*}

After receiving the cores, the T1 cells were removed and the cores were immediately covered by a layer of paraffin wax. This protection from drying remained during the following trimming process. Using an air-cooled diamond coring bit, cylinders of 38 mm diameter were trimmed perpendicular \hl{and parallel} to the bedding plane. The cylinders \hz{of 80 mm height} were also covered by a paraffin layer, before being cut into cylinders of \hz{76 mm height for triaxial tests} or disks \hz{of 10 mm height for isotropic tests} (Fig. \ref{fig:el:straingages})\hz{, using} a diamond wire saw. For protection during storage, the specimens were enveloped by an aluminium foil, covered by a mixture of 70 \% paraffin wax and 30 \% Vaseline oil. 

We carried out a petrophysical characterization, directly after trimming the specimens, and after several months in storage. \hz{Table} \ref{tab:el:1} shows the measured characteristics, determined on cuttings of the cores. We \hz{obtained the specimen volume through hydrostatic weighting in} in \hl{low-odour} hydrocarbon to determine their volume. The dry density was obtained after oven drying at 105  $^{\circ}$C. For the porosity calculation, a  solid density \hl{$\rho_s=$ 2.69 g/cm$^3$ \mbox{\citep{Conil201861}} was assumed}. A chilled mirror tensiometer (WP4, Decagon brand) provided measurements of the suction $s$. The relatively high degree of saturation \hz{measured after opening the T1 cells indicates a good drying protection within the cells.} \hz{In addition, suction and saturation degree were measured on cuttings which were protected and stored in the same way as the specimens. No significant change of these properties was observed} during storage, \hz{which} confirms the \hl{effectiveness of the} adopted protection methods \hz{and a good preservation of the natural water content.}
\begin{table}
\caption{Mean and standard deviation (in \hl{parenthes\hz{e}s}) of petrophysical measurements done on cuttings of three COx cores; with wet and dry density $\rho$ and $\rho_d$, respectively, porosity $\phi_0$, water content $w$, saturation degree $S_r$ and suction $s$.}
\label{tab:el:1}       % Give a unique label
\begin{tabular}{ lcccccc }
	\hline\noalign{\smallskip}
  	\multirow{2}{*}{Core}	& {$\rho$} & {$\rho_d$ } & {$\phi_0$} & {$w$} & {$S_r$}	& {$s$}\\
  							& $[\text{g/cm}^3]$ & $[\text{g/cm}^3]$  & $[\%]$ & $[\%]$ & $[\%]$	& $[\text{MPa}]$\\
	\noalign{\smallskip}\hline\noalign{\smallskip}
 	{EST} 		& 2.37 		& 2.22 		& 17.9 		& 7.5 		& 92.5 		& 24.2\\
 	{53650}		& (0.00) 	& (0.01) 	& (0.2)  	& (0.1) 	& (0.8)		& (2.1)\\
 	\noalign{\smallskip}\hline\noalign{\smallskip} 
 	{EST} 		& 2.38 		& 2.21 		& 18.2 		& 7.9 		& 95.3 		& 17.4 \\
 	{57185}		& (0.00) 	& (0.00) 	& (0.2)  	& (0.1) 	& (0.7)		& (0.1)	\\
 	\noalign{\smallskip}\hline\noalign{\smallskip} 
 	{EST} 		& 2.39 		& 2.22 		& 17.9 		& 7.8 		& 96.3 		& 22.8 \\
 	{58132}		& (0.02) 	& (0.02) 	& (0.6)  	& (0.2) 	& (3.6)		& (1.3)	\\
\noalign{\smallskip}\hline
\end{tabular}
\end{table}

\subsection{Isotropic compression cell}
\label{sec:el:3.2}
\begin{figure}[htbp]
  \includegraphics[width=\columnwidth]{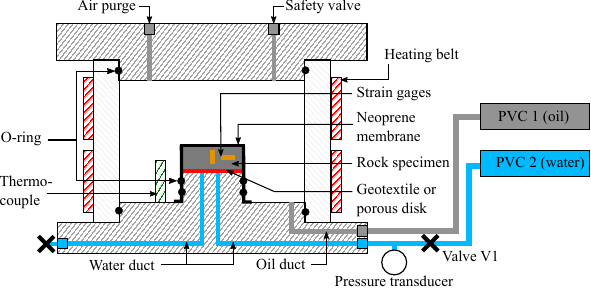}
\caption{Temperature controlled high pressure isotropic cell, accommodating a rock \hl{specimen} equipped with strain gages.}
\label{fig:el:cell}       % Give a unique label
\end{figure}

For the laboratory investigations, we used a high-pressure isotropic thermal compression cell, which is presented in Fig. \ref{fig:el:cell} \citep{Tang200845,Mohajerani201211,Belmokhtar201722,Belmokhtar201787,Braun2019}. This cell allows \hl{us} to test specimens with 38 mm diameter and variable height, placed inside a neoprene membrane in the center of the cell. A \hl{silicone} heating belt \hl{is wrapped around the isotropic cell, which is able to heat up the device. The belt is} coupled with a temperature sensor \hl{in the vicinity of} the specimen, \hl{which allows us to \hz{maintain}} a constant cell temperature \hl{or impose heating/cooling under constant rates}. The cell is filled with silicone oil, which can be set under a pressure of up to 40 MPa by a pressure volume controller (PVC 1, GDS brand). The neoprene membrane isolates the specimen from the silicone oil, so that the pressure of the pore fluid can be controlled independently. The pore pressure is applied from the bottom of the specimen through a porous disk connected to the pressure volume controller PVC 2. During the experiments, we measure the specimen strains locally, using an axial and a radial strain gage (Kyowa brand), glued on the surface of the specimen (Fig. \ref{fig:el:straingages}).

\subsection{Triaxial compression cell}
\label{sec:el:3.2}
\begin{figure}[htbp]
  \includegraphics[width=\columnwidth]{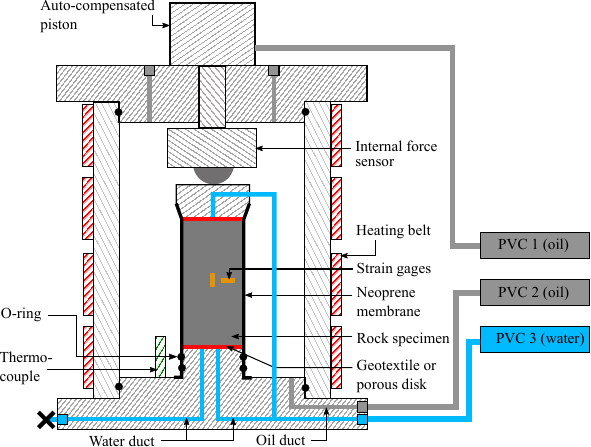}
\caption{Temperature controlled high pressure triaxial cell with an auto-compensated load piston and strain gage measuring system.}
\label{fig:el:triax}       % Give a unique label
\end{figure}

A conventional rock mechanics high pressure triaxial cell was used for experiments with deviatoric loading (Fig. \ref{fig:el:triax}). A triaxial cell is indispensable for determining the Young moduli and Poisson ratios, necessary for a \hl{complete} transverse isotropic parameter set. The device used is similar to the isotropic cell, \hl{including} a pressurized \hl{chamber connected to} PVC 2, which applies confining pressure to the cylindrical specimen (38 mm diameter, variable height). The specimen, isolated by a neoprene membrane, is submitted to pore pressures through porous stones on its top and bottom surfaces, controlled by PVC 3. Specimen deformations are detected by strain gages attached at specimen mid-height (Fig. \ref{fig:el:straingages}). \hl{The same heating system as in the isotropic cell was used for this device, with a silicone} heating belt, connected to an internal thermocouple \hl{to regulate} the cell temperature. The main difference to the isotropic cell is the auto-compensated piston, which allows \hl{us} to apply axial \hl{deviatoric} loads. \hl{An} auto-compensation \hl{chamber pressurized by the confining fluid $\sigma_{\text{rad}}$ (situated in the piston housing, not shown in Fig. \ref{fig:el:triax})} keeps the piston in equilibrium, when the pressure applied by PVC 1 is zero. \hl{In this case}, the specimen is in an isotropic stress state \hl{where both axial and radial total stresses are equal ($\sigma_{\text{ax}}=\sigma_{\text{rad}}$). The pressure applied by PVC 1 controls then only the additional deviatoric part of the stress $q=\sigma_{\text{ax}}-\sigma_{\text{rad}}$ (e.g. \mbox{\citealp{Fortin2005}}).}

\subsection{Testing Procedure}
\label{sec:el:3.3}
\begin{figure}[htbp]
\includegraphics[width=\columnwidth]{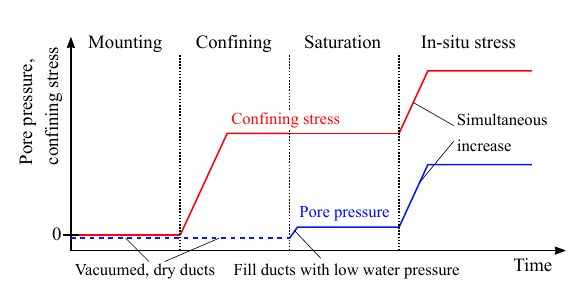}
\caption{\hl{Mounting and saturation procedure for COx specimens. Specimens are first mounted with dry water-ducts. Vacuum is applied to the ducts in order to evacuate air. The isotropic confining stress is then increased and kept constant. When monitored strains are stable, the water saturation is started with a low back pressure to limit poroelastic strains. After the hydration swelling stabilized, both pore pressure and confining stress are increased simultaneously until the desired stress level close to the in-situ one is reached. Again, strains are monitored and when the deformations are in equilibrium, further testing can be started.}}
\label{fig:el:saturation_procedure}       % Give a unique label
\end{figure}

The COx specimens were mounted in the isotropic and triaxial cells in their initial state at 25 $^{\circ}$C. \hl{An initial sequence of  consolidation, saturation and in-situ stress application was followed, shown schematically in Fig. \ref{fig:el:saturation_procedure}.} 

\hl{First, the specimens were consolidated at constant water content under isotropic stress with a loading rate of 0.1 MPa/min, while the drainage system was kept dry.} Most of the specimens were brought to a confining pressure close to the in-situ effective stress between 8 MPa and 10 MPa. \hz{Three} specimens were \hl{brought to a} higher confining stress to investigate the effect of confining \hl{pressure} on swelling strains during \hl{the following} hydration. An overview of all tested specimens with their respective initial confining stress is given in Table \ref{tab:el:overview}. 

\begin{table*}
\caption{Overview of tested specimens with their initial saturation phase under stress and their respective subsequent loading paths; the inclined arrows ($\searrow$,$\nearrow$) indicate that several load steps were carried out between mentioned effective stresses.}
\begin{tabular}{llcccc}
\hline\noalign{\smallskip}
\multirow{2}{*}{\hz{Specimen}} & \multirow{2}{*}{Core} & \multicolumn{2}{c}{Saturation phase} & \multicolumn{2}{c}{Loading path}              \\
                        &                       & Confining stress      & Volumetric swelling     & Isotropic tests:	& Deviatoric tests: \\
\#                      & EST                   & {[}MPa{]}             & {[}\%{]}     &  $\sigma'_{\mathrm{rad}}=\sigma'_{\mathrm{ax}}$ {[}MPa{]}               & $q$ {[}MPa{]}           \\ \noalign{\smallskip}\hline\noalign{\smallskip}
ISO1                    & 57185                 & 10                    & 0.71         & 10$\searrow$1$\nearrow$18                 & -                   \\
ISO2                    & 53650                 & 8                     & 0.86         & -                       & -                   \\
ISO3                    & 53650                 & 8                     & 0.80         & -                       & -                   \\
ISO4                    & 53650                 & 8                     & 0.72         & 8$\nearrow$15$\searrow$2                  & -                   \\
ISO5                    & 53650                 & 10                    & 0.61         & 10$\nearrow$36$\searrow$0.5               & -                   \\
ISO6                    & 53650                 & 14                    & 0.57         & 14$\leftrightarrow$16                   & -                   \\
ISO7                    & 53650                 & 10                    & 0.66         & -                       & -                   \\
ISO8                    & 53650                 & 8                     & 0.95         & -                       & -                   \\
ISO9                    & 53650                 & 30                    & 0.15         & -                       & -                   \\
ISO10                   & 58132                 & 20                    & 0.33         & 20$\leftrightarrow$18                       & -                   \\
DEV1                    & 57185                 & 10                    & 0.55         & 30$\rightarrow$20$\rightarrow$10                & 0$\leftrightarrow$5 (@$\sigma'_{\mathrm{rad}}=$ 30/20/10)                \\
DEV2                    & 57185                 & 10                    & 0.62         & -                       & 0$\leftrightarrow$5 (@$\sigma'_{\mathrm{rad}}=$ 11) \\
DEV3                    & 57185                 & 20                    & 0.42         & -                   & 0$\leftrightarrow$10 (@$\sigma'_{\mathrm{rad}}=$ 10)               \\ \noalign{\smallskip}\hline
\end{tabular}
\label{tab:el:overview} 
\end{table*}

\begin{figure}[htbp]
  \includegraphics[width=\columnwidth]{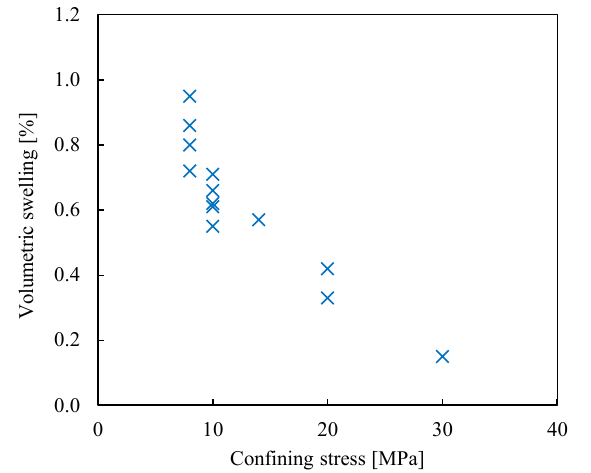}
\caption{Measured swelling strains of specimens during hydration under \hl{100 kPa pore pressure and} various confining stresses.}
\label{fig:el:swellingstress}       % Give a unique label
\end{figure}

When specimen deformations remained stable after application of the confining stress, \hl{we assume hydraulic equilibrium and a complete dissipation of any generated over-pressure in the pores during the loading, similar to the end of secondary consolidation in soils. The} air in the dry drainage ducts was evacuated by vacuum \hl{for several minutes and the} ducts were then filled with synthetic pore water under a pressure of 100 kPa. To prepare synthetic water, salts were added to \hl{de-mineralized} water according to a recipe provided by Andra, to obtain a fluid composition close to the in-situ one. \hz{When injecting pore fluid into a initially partially unsaturated specimen, we expect a total volume increase composed of i) the volume changes of an unsaturated medium due to the reduction of suction and the adsorption of water and   ii) the poroelastic response of a saturated medium due to pore pressure increase. The former occurs, when the pore pressure increases from a negative value (corresponding to suction) to zero, while the latter happens under positive pore pressure changes, once the medium is saturated. In order to better distinguish these effects in the experiments, and generate first a predominantly unsaturated response, the initially applied water pressure was chosen relatively small. In this way, we try to limit \hl{poroelastic deformations due to pore pressure changes} and observe hydration strains only. A certain positive pressure is however required to fill and saturate the drainage lines of the device.} Hydration resulted in swelling strains that stabilized after about \hz{five} days on all \hl{specimens}, showing similar transient characteristics. Tab. \ref{tab:el:overview} lists the measured volumetric swelling strains, which decrease with increasing confining stress (Fig. \ref{fig:el:swellingstress}). \hl{Here we neglect however possible size effects due to different specimen dimensions and a potentially incomplete saturation at 100 kPa pore pressure.}

After hydration, pore pressure and confining pressure were increased simultaneously \hl{by} 4 MPa under a constant Terzaghi effective stress. The applied pressures were increased until the pore pressure reached 4 MPa. Consequently, the \hl{specimens} reached Terzaghi effective stresses, which were equal to the initial confining stress \hz{in the saturation phase} described in Table \ref{tab:el:overview}. 
\hl{We can assume complete saturation when the recorded strains stabilized, indicating hydraulic equilibrium. \mbox{\cite{Rad1984}} conducted calculations based on Boyle's and Henry's laws to estimate the necessary back-pressure for a complete dissolution of the air trapped in pores. They showed that for saturating a specimen which is initially saturated at 90 \% and submitted to a vacuum of 80 kPa, one requires around 150 kPa back pressure. Here, while having the same saturation degree and imposed vacuum, we applied a much higher pore pressure of 4 MPa. \mbox{\cite{Favero2018}} carried out tests of the Skempton coefficient of initially unsaturated Opalinus clay at different pore pressure levels. For pore pressures higher than 2 MPa, they found that the Skempton coefficient remained stable, which indicated complete saturation. 
%Another evidence for a complete saturation of the specimens tested in this study is the inter-compatibility of the measured poroelastic parameters, presented in the following sections. We determined the coefficients within a saturated poroelasticity framework, which would create compatibility issues if the material was unsaturated.
Note that during this last saturation phase, no distinction of the measured strains between poroelastic deformations and possible additional swelling (due to further saturation by the increasing back-pressure) could be made.}

\begin{figure*}[htbp]
  \includegraphics[width=\textwidth]{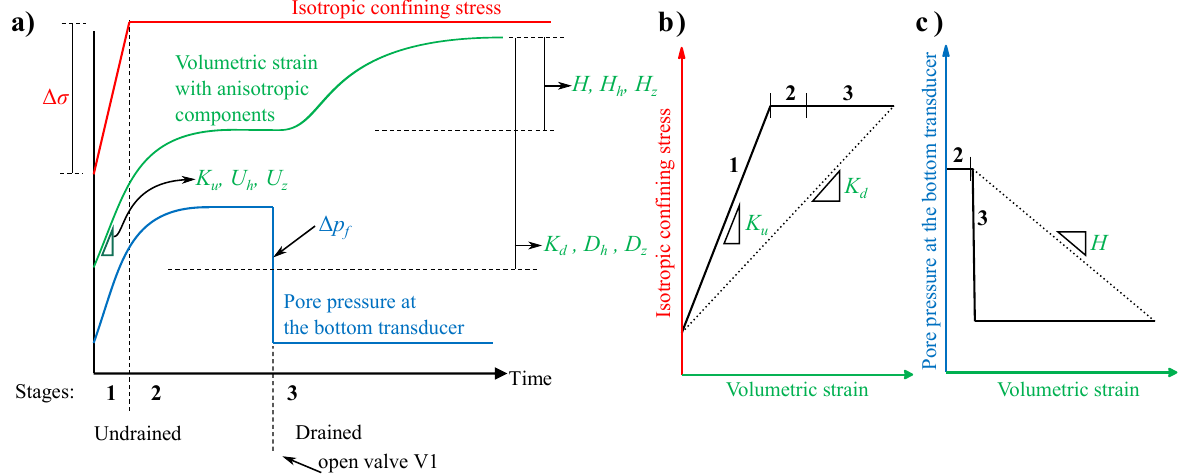}
\caption{\hl{Schematic representation of the step testing procedure, adopted from \mbox{\cite{Braun2019}} and \mbox{\cite{Hart2001}}. \hz{a) Applied and measured changes of stress, pore pressure and strain with time.} A rapid increase of the confining stress $\Delta\sigma$ allows us to measure the undrained properties $K_u$ (Eq. (\ref{eq:el:3})), $U_h$ and $U_z$ (Eq. (\ref{eq:el:Udef})) from the initial slope of the \hz{stress-}strain response, \hz{shown in b)}. 
The drainage system is closed in this phase. After pore pressure and strain remain constant, the drainage system is opened and a change of pore pressure $\Delta p_f$ is imposed. This pore pressure change creates a strain response, providing us the (secant) Biot modulus $H$ \hz{in c) (see also} Eq. (\ref{eq:el:2})), with $H_h$ and $H_z$ (Eq. (\ref{eq:el:Hdef})). The drained parameters $K_d$ (Eq. (\ref{eq:el:1})), $D_h$ and $D_z$ (Eq. (\ref{eq:el:Ddef})) are measured from the secant of the strain response in drained state, before and after the test, \hz{as shown in b)}.}}
\label{fig:el:procedure}       % Give a unique label
\end{figure*}

\hl{A step testing procedure, presented by \mbox{\cite{Braun2019}}, was then utilized to determine the poroelastic parameters of the material in isotropic tests. This procedure (Fig. \ref{fig:el:procedure}) consists of increasing or decreasing the confining pressure rapidly at a rate of 0.1 MPa/min and keeping it then constant at a certain level ($\Delta\sigma$). This generates first an undrained response, which permits evaluation of the undrained bulk modulus $K_u$ (Eq. (\ref{eq:el:3})) as a tangent modulus, together with the anisotropic moduli $U_h$ and $U_z$ (Eq. (\ref{eq:el:Udef})). During a constant load phase, the pore pressures generated in the first phase are allowed to drain by releasing the undrained pore pressure changes through the drainage valve. A drained response can be observed, when all pore pressures have dissipated and the deformations stabilize, providing the drained bulk modulus $K_d$ (Eq. (\ref{eq:el:1})) with $D_h$ and $D_z$ (Eq. (\ref{eq:el:Ddef})). When the confining pressure is held constant and the imposed pore pressure is changed instantly ($\Delta p_f$) by the PVC, a transient deformation can be observed. After stabilization of these deformations, the Biot modulus $H$ (Eq. (\ref{eq:el:2})) with $H_h$ and $H_z$ (Eq. (\ref{eq:el:Hdef})) can be evaluated. This pore pressure loading step can also be carried out independently of a precedent confining pressure loading.}

\begin{figure*}[htbp]
  \includegraphics[width=\textwidth]{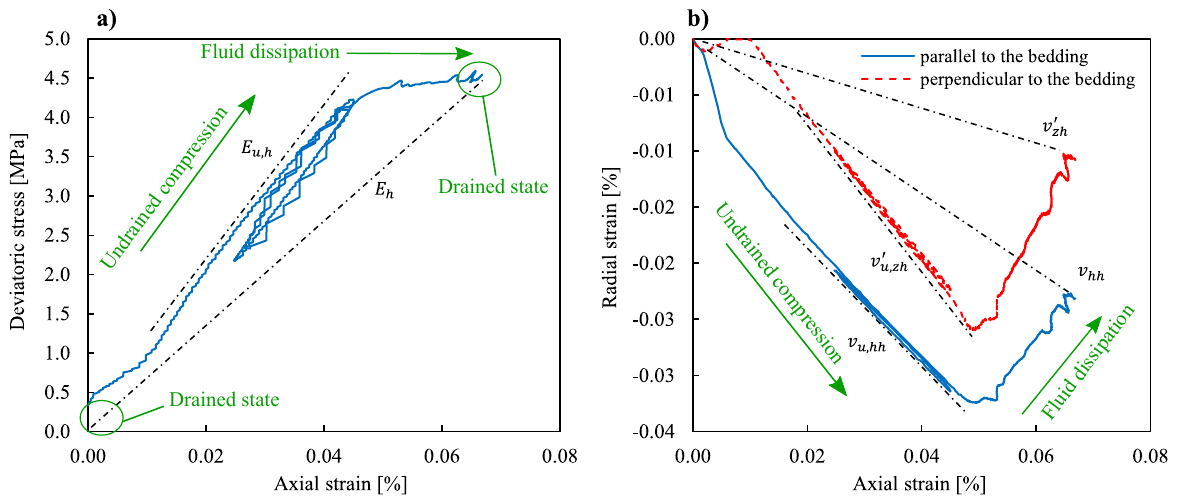}
\caption{Measured deformations during a deviatoric test on \hz{specimen} DEV2 loaded parallel to the bedding plane in a) function of deviatoric stress $q$, highlighting the determination of the Young moduli. \hl{The fast loading shows a linear undrained response, while pore pressure diffusion induces further compression with time, once the deviatoric stress is held constant. Note that a fast unloading reloading cycle was only carried out for this specific \hz{specimen}. No drainage effects in form of a hysteresis were found during the cycle, evidencing a constant undrained condition. b) Radial strains with respect to axial strains indicating the Poisson ratios, also describing a clear difference between undrained response under fast loading and time dependent fluid dissipation, until reaching drained condition.}}
\label{fig:el:stressstrain_example}       % Give a unique label
\end{figure*}

\hl{In the deviatoric tests we proceeded \hz{similarly} to the isotropic loading tests, by applying the deviatoric load $q$ in a relatively rapid loading of 0.1 MPa/min, where we assume undrained conditions. In this phase, we determine the undrained coefficients $E_{u,i}$ and $\nu_{u,i}$ (see also Sec. \ref{sec:el:2} for the parameter definitions and their evaluation) from the initial slopes of the measurements of stress and strain changes (Fig. \ref{fig:el:stressstrain_example}). 
Once the target deviatoric stress is reached, we keep it constant, causing a fluid dissipation and equilibration of the pore pressure field inside the specimen. We \hz{assume} completed drainage when measured strains stabilize. Doing so, we can assure drained conditions before and after each step loading. One can evaluate the drained Young's moduli and Poisson's ratios as secant parameters, comparing strains and stresses in the drained states before and after the step.
Deviatoric tests on specimens DEV1 and DEV3 were conducted perpendicular to the bedding plane, providing $E_{z}$, $\nu_{zh}$, $E_{u,z}$ and $\nu_{u,zh}$ (compare Fig. \ref{fig:el:straingages}), whereas the test parallel to the bedding plane on DEV2 provided $E_{h}$, $\nu_{hh}$, $\nu_{zh}'$, $E_{u,h}$, $\nu_{u,hh}$ and $\nu_{u,zh}'$. }

\begin{figure*}[htbp]
  \includegraphics[width=\textwidth]{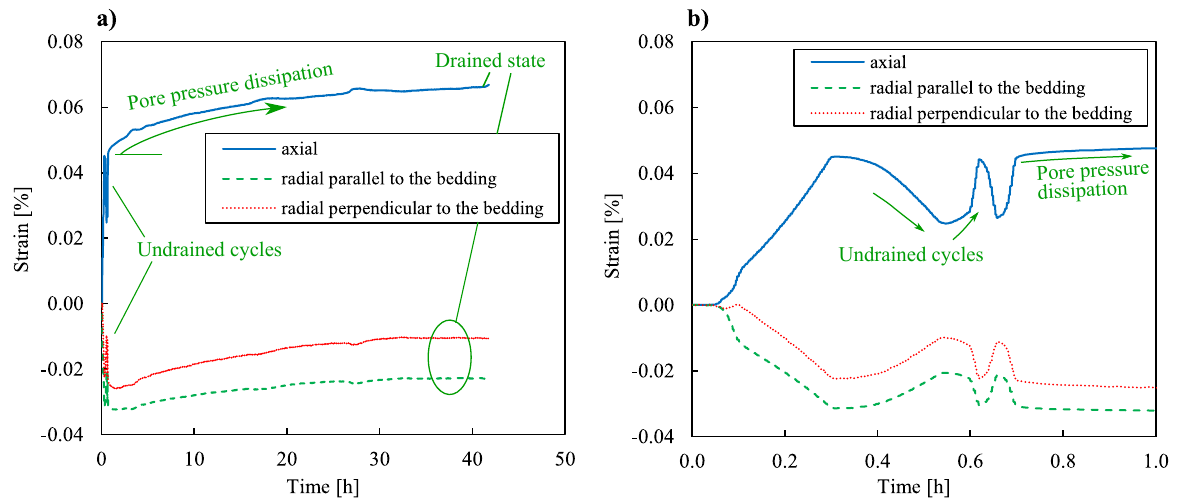}
\caption{\hl{Measured deformations during a deviatoric test on \hz{specimen} DEV2 loaded parallel to the bedding plane in a) function of time, with b) zoom on the first hour of a).}}
\label{fig:el:strain_example}       % Give a unique label
\end{figure*}

\hl{Undrained conditions during the step increase of $q$ were verified in a single, quick unloading/reloading cycle on specimen DEV2 (Fig. \ref{fig:el:strain_example}). We observe a constant stress-strain slope with no hysteresis after the change of loading direction (Fig. \ref{fig:el:stressstrain_example}a). In parallel, constant slopes indicating constant $\nu_{u,i}$ are observed in Fig. \ref{fig:el:stressstrain_example}b. In a perfectly poroelastic material, a hysteresis can only appear due to time dependent fluid diffusion processes. If the loading is fast enough, no diffusion occurs, as the specimen remains in constant undrained conditions. In the same test, after the rapid loading (Fig. \ref{fig:el:stressstrain_example}, \ref{fig:el:strain_example}), one can see the additional fluid-dissipation-induced strains with respect to time, as a transient between undrained and drained state.}

\begin{figure}[htbp]
  \includegraphics[width=\columnwidth]{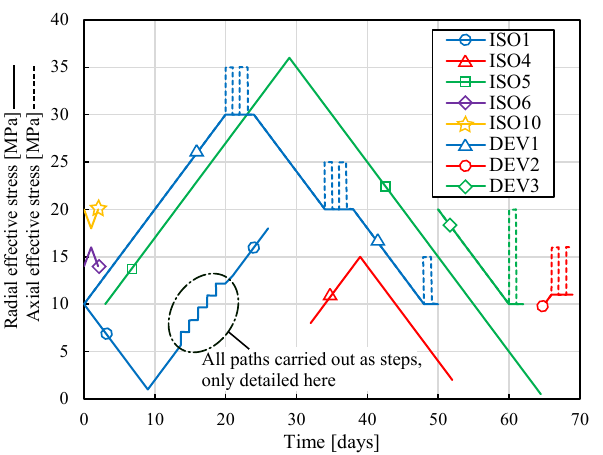}
\caption{\hl{Schematic representation of the loading paths carried out in this work. The ISO specimens were loaded in steps by changing isotropic confining stress or pore pressure. Note that the inclined lines represent several step loadings, which were not detailed here. Also the DEV specimens were submitted isotropic step loading (solid line), and to additional deviatoric loadings (dashed line) in the form of step-cycles of axial stress. The constant load phase after each isotropic or deviatoric step lasted for at least one day, in order to attain a pore pressure equilibrium. For a better overview, the loading paths were assembled with offsets in the time axis.}}
\label{fig:el:paths}       % Give a unique label
\end{figure}

\hl{For isotropic tests \hz{($\sigma'=\sigma'_{\mathrm{ax}}=\sigma'_{\mathrm{rad}}$)}, the measured parameters are hereafter presented with respect to the mean between initial and final effective stress applied during each step test, \hz{i.e. a parameter measured under a step increase from 4.0 to 6.0 MPa is shown as a data-point at 5.0 MPa.} All discussed coefficients were measured under unloading or reloading and are therefore supposed to be elastic. Several subsequent step tests were carried out, starting from the initial stress level and then gradually increasing or decreasing the effective stress in each step with an increment of between 1.0 and 5.0 MPa stress (Fig. \ref{fig:el:paths}). These ramps are indicated in Tab. \ref{tab:el:overview} with inclined arrow symbols ($\searrow$,$\nearrow$). Horizontal arrows ($\rightarrow$) characterise only one step between the indicated effective stresses, which was repeated in the case of a double arrow ($\leftrightarrow$). The same applies for the deviatoric tests on specimens DEV1 - DEV3. Specimen DEV1 was loaded and unloaded by applying a devatoric load $q$ between 0 and 5 MPa, which was repeated at effective radial stresses $\sigma'_{\mathrm{rad}}$ of 30, 20 and 10 MPa. \hz{Specimens} DEV2 and DEV3 were only tested at 11 MPa and 10 MPa effective radial stress, respectively. The resulting parameters are plotted with respect to the average of the inital and final isotropic effective stress $\sigma'$ at each step loading.
Specimens without any loading path \hz{given in Table \ref{tab:el:overview}} were not tested under hydro-mechanical experiments. They were either only tested in thermal tests (not discussed here) or had to be abandoned due to technical problems, and are listed only for their data on the swelling upon resaturation. }

\section{Experimental results}
\label{sec:el:4}
\subsection{Stress dependent behaviour under isotropic loading}
\label{sec:el:4.2}
\begin{figure}[htbp]
  \includegraphics[width=\columnwidth]{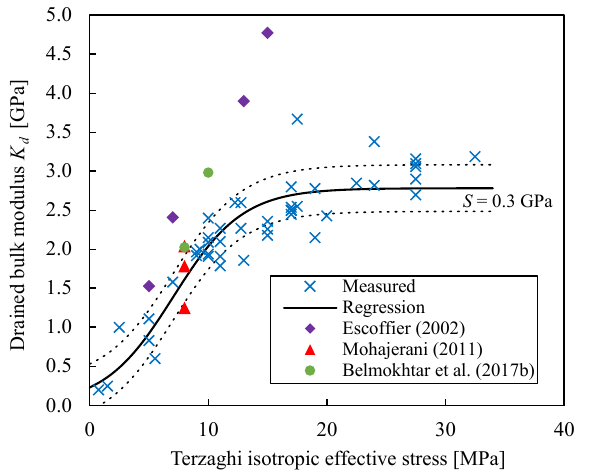}
\caption{Measured stress dependent drained bulk modulus $K_d$ \hl{compared with a fitted parameter set (Sec. \ref{sec:el:5}). $S$ \hz{and dotted lines} indicate the standard deviation of the regression.}}
\label{fig:el:Kdsig}       % Give a unique label
\end{figure}

\begin{figure}[htbp]
  \includegraphics[width=\columnwidth]{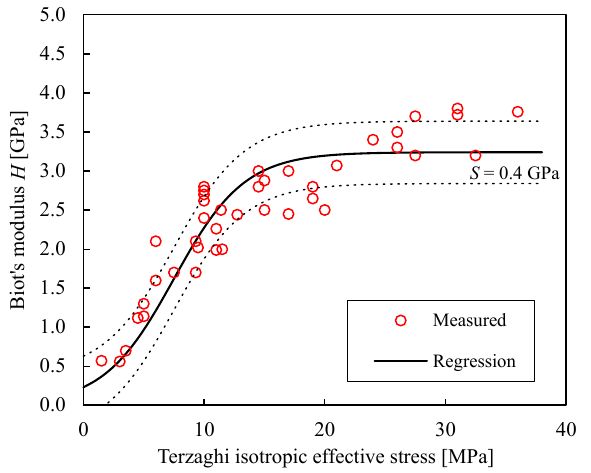}
\caption{Measured stress dependent Biot's modulus $H$ \hl{compared with a fitted parameter set (Sec. \ref{sec:el:5}). \hz{$S$ and dotted lines indicate the standard deviation of the regression.}}}
\label{fig:el:Hsig}       % Give a unique label
\end{figure}

%
%As there is no apparent temperature dependency of elastic properties, we do not differ in the following between measurements of different temperature levels.
\hl{The conducted} isotropic hydromechanical tests provided measurements of the moduli $K_d$ and $H$ with a strong dependency on the effective stress, \hl{consistent} for all tested specimens (Fig. \ref{fig:el:Kdsig} and \ref{fig:el:Hsig}). 
\cite{Zimmerman198612} showed, that under the condition of a constant modulus $K_s$, a stress dependency of the tangent bulk moduli of an isotropic material can be expressed \hl{as a} function of the Terzaghi \hl{mean} effective stress. In Section \ref{sec:el:5} we demonstrate, that assuming a constant $K_s$ we are indeed able to reproduce the measured COx properties. \hl{As} the COx claystone is not isotropic, we decided here to present the following measurements with respect to the Terzaghi \hl{isotropic} effective stress. \hl{The evaluated parameters are later analysed in Sec. \ref{sec:el:5} as representative functions of the isotropic effective stress.}

The moduli \hl{$K_d$ and $H$} are smaller than 1.0 GPa for low Terzaghi effective stresses and increase with increasing effective stress \hl{(Fig. \ref{fig:el:Kdsig} and \ref{fig:el:Hsig}).} With an effective stress larger than \hl{around} 20 MPa, the moduli appear to remain constant\hl{, with values of around 3.0 and 3.5 GPa, respectively.}
For $K_d$ \hl{at} effective stresses around 10 MPa, we can observe a good \hl{agreement} with the findings of \cite{Mohajerani2011} and \cite{Belmokhtar201787}. The modulus $H$ was previously  measured by \cite{Belmokhtar201787} at around 10 MPa effective stress with 3.47 and 2.24 GPa, \hl{somewhat higher than the present findings.}

\begin{figure}[htbp]
  \includegraphics[width=\columnwidth]{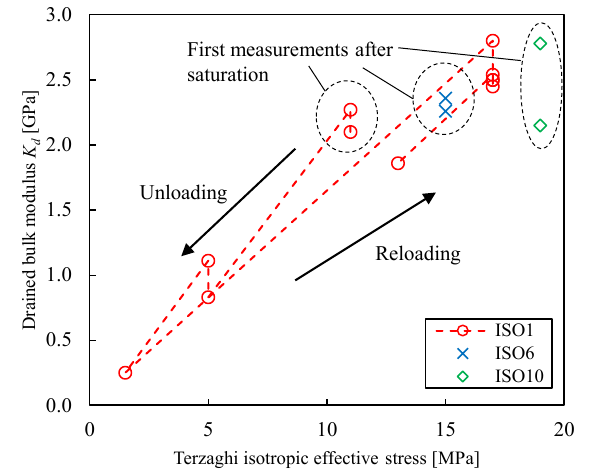}
\caption{Measured stress dependent drained bulk modulus $K_d$ on \hz{specimens} ISO1, ISO6 and ISO10, illustrating the reversible reduction of stiffness with decreasing effective stress.}
\label{fig:el:damage}       % Give a unique label
\end{figure}

Interestingly, the reduction of stiffness \hl{with effective isotropic stress} appears to be reversible, as can be seen in Fig. \ref{fig:el:damage}. Here we compare the drained bulk modulus, measured on \hz{specimen} ISO1 after subsequent unloading steps, starting from the initial under-stress-saturated state. The bulk modulus decreases through unloading. After reloading, the drained bulk modulus \hl{comes} back to the same level as before unloading, comparable also to the values on ISO6 and ISO10, for which the effective stress level was not significantly reduced after saturation.

\begin{figure}[htbp]
  \includegraphics[width=\columnwidth]{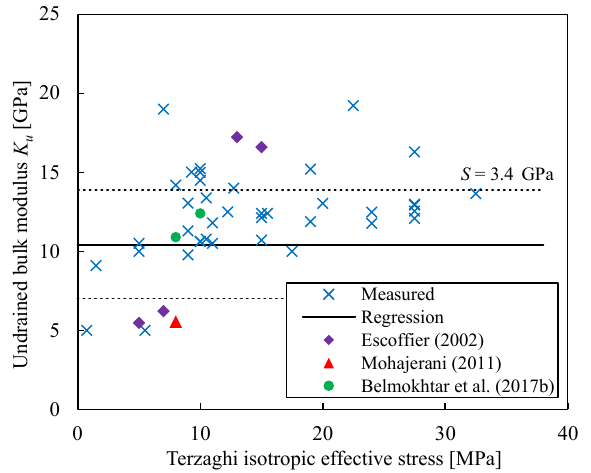}
\caption{Measured undrained bulk modulus $K_u$, \hl{compared with a fitted parameter set (Sec. \ref{sec:el:5}). \hz{$S$ and dotted lines indicate the standard deviation of the regression.}}}
\label{fig:el:Ku}       % Give a unique label
\end{figure}

\hl{For the undrained bulk modulus $K_u$ it is difficult to identify} a stress dependency (Fig. \ref{fig:el:Ku}), \hl{due to} the rather large dispersion of observed values, which was also evidenced in \hl{previous studies \mbox{\citep{Escoffier2002,Mohajerani2011,Belmokhtar201787}}.}

\begin{figure}[htbp]
  \includegraphics[width=\columnwidth]{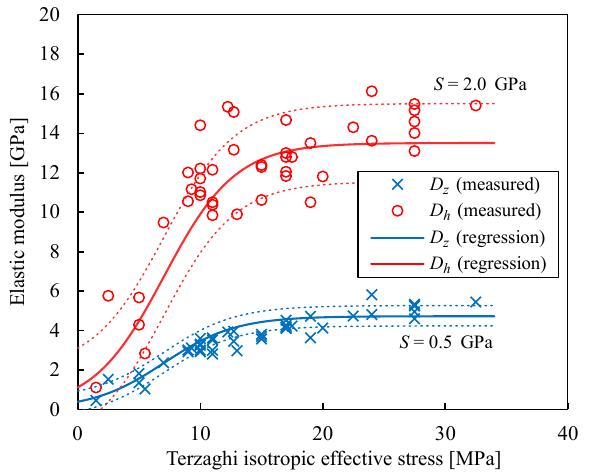}
\caption{Measured anisotropic responses $D_i$ during drained compression, \hl{compared with a fitted parameter set (Sec. \ref{sec:el:5}). \hz{$S$ and dotted lines indicate the standard deviation of the regression.}}}
\label{fig:el:Daniso}       % Give a unique label
\end{figure}

\begin{figure}[htbp]
  \includegraphics[width=\columnwidth]{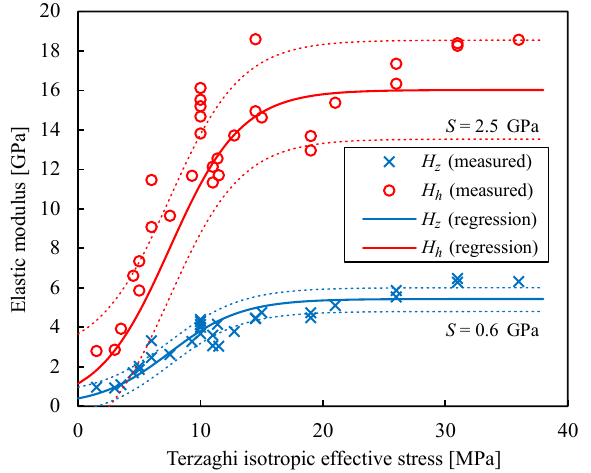}
\caption{Measured anisotropic responses $H_i$ during pore pressure tests, \hl{compared with a fitted parameter set (Sec. \ref{sec:el:5}). \hz{$S$ and dotted lines indicate the standard deviation of the regression.}}}
\label{fig:el:Haniso}       % Give a unique label
\end{figure}

\begin{figure}[htbp]
  \includegraphics[width=\columnwidth]{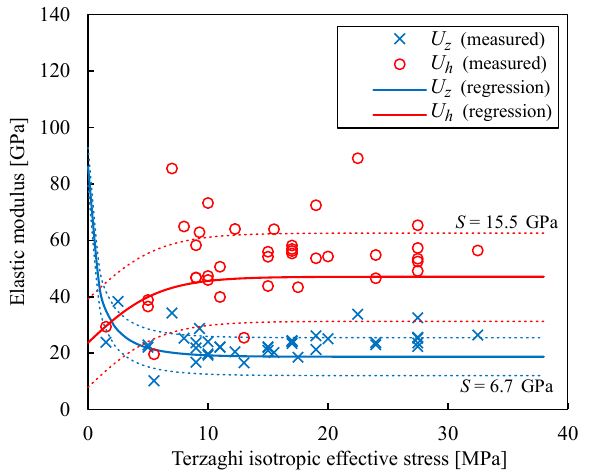}
\caption{Measured anisotropic responses $U_i$ during undrained compression, compared with our best-fit parameter set (Sec. \ref{sec:el:5}). \hz{$S$ and dotted lines indicate the standard deviation of the regression.}}
\label{fig:el:Kuaniso}       % Give a unique label
\end{figure}

The transversely isotropic strain response was recorded simultaneously with $K_d$, $H$ and $K_u$ in all tests, \hl{resulting in the parameters $D_i$, $H_i$ and $U_i$.} (Fig. \ref{fig:el:Daniso}, \ref{fig:el:Haniso} and \ref{fig:el:Kuaniso}). For drained tests ($D_i$, $H_i$) \hl{above 20 MPa effective stress,} one observes axial strains perpendicular to the bedding plane about three times larger than radial ones. In undrained tests, the ratio of anisotropy varied more \hl{significantly} with effective stress, with values around 1.5 for low effective stress and values around 2.0 for 30 MPa effective stress. \hl{Again, the rather large scatter with some outliers of the measured undrained properties becomes visible on the parameters $U_i$ (Fig. \ref{fig:el:Kuaniso}).} 

\subsection{Stress dependent behaviour under deviatoric loading}
\label{sec:el:4.3}
\begin{figure}[htbp]
  \includegraphics[width=\columnwidth]{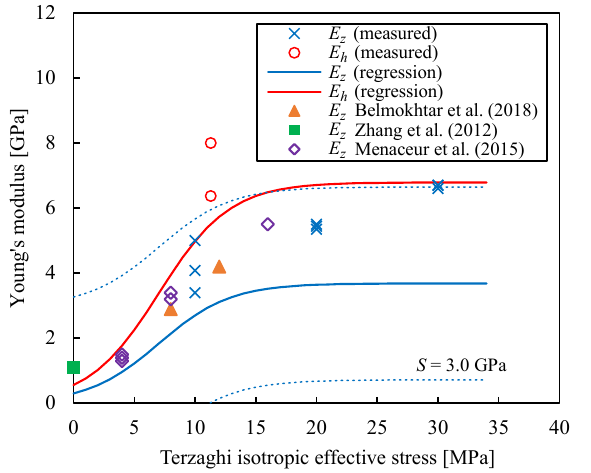}
\caption{Measured stress dependent drained Young's modulus $E_z$ perpendicular and $E_h$ parallel to the bedding plane, \hl{compared with a fitted parameter set (Sec. \ref{sec:el:5}). \hz{$S$ and dotted lines indicate the standard deviation of the regression.}}}
\label{fig:el:Edrained}       % Give a unique label
\end{figure}

\begin{figure}[htbp]
  \includegraphics[width=\columnwidth]{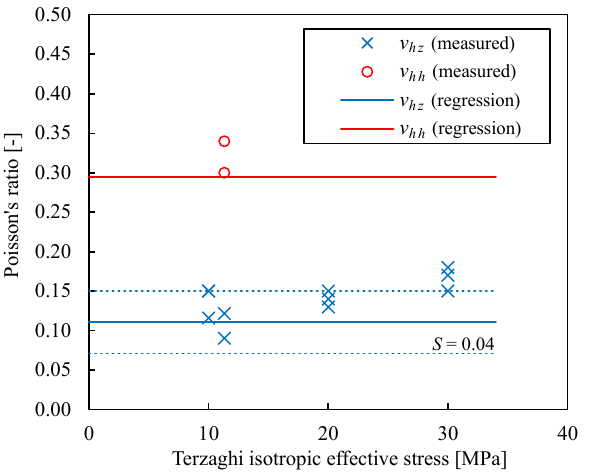}
\caption{Measured drained Poisson's ratio $\nu_{zh}$ perpendicular and $\nu_{hh}$ parallel to the bedding planes with respect to effective stress, \hl{compared with a fitted parameter set (Sec. \ref{sec:el:5}). \hz{$S$ and dotted lines indicate the standard deviation of the regression.}}}
\label{fig:el:vdrained}       % Give a unique label
\end{figure}

In the triaxial tests, we measured $E_{z}$ and $\nu_{zh}$ on specimens DEV1 and DEV3 for effective confining stresses from 10 to 30 MPa \hl{(Fig. \ref{fig:el:Edrained}, \ref{fig:el:vdrained}).} Whereas the Poisson ratio remained more or less constant with values between 0.10 and 0.17, the Young modulus $E_{z}$ increased with increasing confining stress from around 3.4 to 6.7 GPa. \hl{Compared with the values of $E_{z}$ from \mbox{\cite{Zhang201279}}, \mbox{\cite{Menaceur201529}} and \mbox{\cite{Belmokhtar201819}}, a similar stress dependency can be observed.}
\hl{The parameters $E_{h}$ and $\nu_{hh}$ on \hz{specimen} DEV2 were measured only for 10 MPa isotropic effective stress, with values for $E_{h}$ between 6.4 and 8.0 GPa and $\nu_{hh}$ between 0.30 and 0.34. Note that only two measurements with a relatively large difference were taken for this loading direction.}

\begin{figure}[htbp]
  \includegraphics[width=\columnwidth]{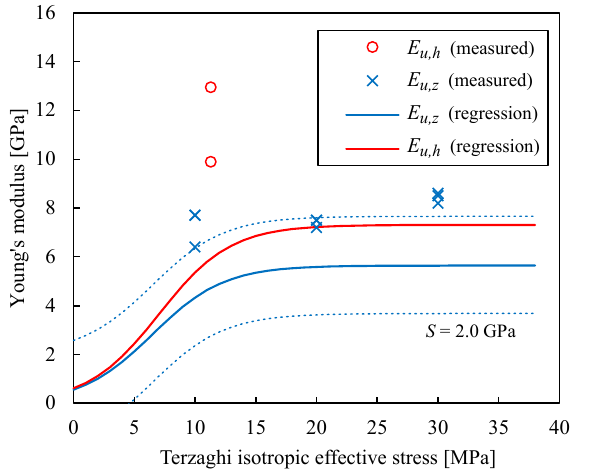}
\caption{Measured stress dependent undrained Young's modulus $E_{u,z}$ perpendicular and $E_{u,h}$ parallel to the bedding plane, \hl{compared with a fitted parameter set (Sec. \ref{sec:el:5}). \hz{$S$ and dotted lines indicate the standard deviation of the regression.}}}
\label{fig:el:Eu}       % Give a unique label
\end{figure}

\begin{figure}[htbp]
  \includegraphics[width=\columnwidth]{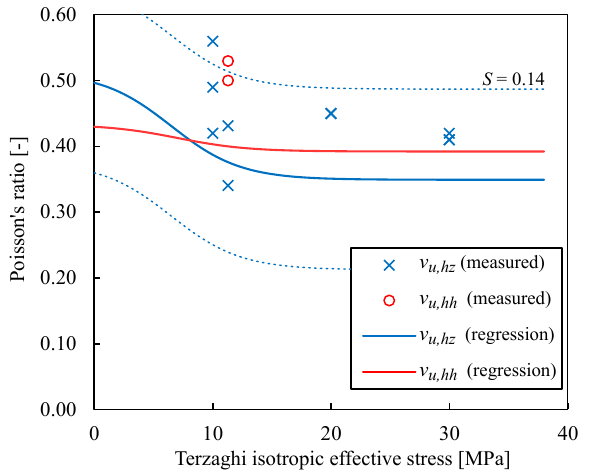}
\caption{Measured undrained Poisson's ratio $\nu_{u,zh}$ perpendicular and $\nu_{u,hh}$ parallel to the bedding planes with respect to effective stress, \hl{compared with a fitted parameter set (Sec. \ref{sec:el:5}). \hz{$S$ and dotted lines indicate the standard deviation of the regression.}}}
\label{fig:el:vu}       % Give a unique label
\end{figure}

%ADD comparison to literature for v other shales
As expected, the undrained response during \hl{triaxial} tests \hz{produced} higher moduli than the drained ones. The parameter $E_{u,z}$ was found between 6.4 and 8.6 GPa and $E_{u,h}$ between 9.8 and 10.8 GPa \hl{(Fig. \ref{fig:el:Eu}).} The anisotropy of the undrained moduli shows a ratio $E_{u,h}/E_{u,z}$ around 1.4, whereas the drained moduli show a ratio $E_{h}/E_{z}$ of around \hl{1.8}. 
\hl{Interestingly, we can observe relatively large undrained Poisson's ratios, with values higher than 0.35 and a significant scatter (Fig. \ref{fig:el:vu}).
In contrast to isotropic elastic materials, the Poisson ratios of anisotropic elastic materials have in theory no bounds \mbox{\citep{Ting2005}}. High Poisson ratios $\nu>$ 0.5 have been observed on rocks in the laboratory (e.g. \citealp{Vutukuri1974,Hatheway1982,Islam2013}), which can be related to significant anisotropy \mbox{\citep{Gercek2007}}.
As for $E_{h}$ and $\nu_{hh}$, note that only two measurements have been made for $E_{u,h}$ and $\nu_{u,hh}$, which show a relatively large difference.}

%ADD comparison to literature for Eu other shales
%ADD comparison to literature for v other shales

\section{Regression analysis}
\label{sec:el:5}

The experimental results evidence an increase of the \hl{moduli $K_d$, $D_i$, $H$, $H_i$ and $E_i$} with increasing Terzaghi effective stress, starting from a certain value at zero effective stress and reaching a plateau at a given effective stress (e.g. Fig. \ref{fig:el:Kdsig}). To describe this trend, we chose \hl{an empirical,} sigmoid function, used regularly in rock mechanics literature (e.g. \citealp{Zimmerman1991,Hassanzadegan201442,Ghabezloo201414}). \hl{Here we attribute the sigmoid function to $E_z$: }
\begin{equation}
\mathcolorbox{
\frac{1}{E_z} = \frac{1}{{{E_z^\infty }}} + \left( {\frac{1}{{{E_z^0}}} - \frac{1}{{{E_z^\infty }}}} \right)\exp \left( { - \beta {\sigma'}} \right)
}
\label{eq:el:Kdmodel}
\end{equation}
where \hl{$E_z$} increases with \hl{the Terzaghi isotropic effective stress} $\sigma'$ from a lower limit \hl{${E_z^0}$} to an upper limit \hl{$E_z^\infty $}, with the shape of transition governed by a parameter $\beta$.
\hl{In addition, we introduce an anisotropy ratio $R_E$, which allows us to calculate $E_h$:}

\begin{equation}
\mathcolorbox{
R_E=E_h/E_z
}
\label{eq:el:R_E}
\end{equation}

\hl{Knowing $\nu_{hz}$ and $\nu_{hh}$, the parameters $K_d$ and $D_i$ can then be evaluated using Eq. (\ref{eq:el:Ei}) and (\ref{eq:el:Di}). With Eq. (\ref{eq:el:2}), one obtains $H$ by inserting the unjacketed compression modulus $K_s$. We use then the ratio $R_H$ as another fitting parameter, which provides us $H_i$ (Eq. (\ref{eq:el:2})). The Biot coefficients $b_i$ can now be calculated with Eq. (\ref{eq:el:bh}) and (\ref{eq:el:bz}) and the elastic stiffness matrix $M_{ij}$ by Eq. (\ref{eq:el:M_matrix}) - (\ref{eq:el:M33}).}

\hl{To evaluate $M$, one can insert Eq. (\ref{eq:el:Kphi}) in Eq. (\ref{eq:el:M}), which eliminates $K_f$ and $K_\phi$ and requires only one additional fitting parameter $K_u$: }

\begin{equation}
\begin{split}
\frac{1}{M}=2 \left( 1-{b_h}\right)  \left( \frac{{b_h} \left( 1-{\nu_{hh}}\right) }{{E_h}}-\frac{{b_z} {\nu_{hz}}}{{E_z}}\right) + \\
\frac{\left( 1-{b_z}\right)  \left( {b_z}-2 {b_h} {\nu_{hz}}\right) }{{E_z}}+\frac{\left( \frac{1}{{K_d}}-\frac{1}{{K_s}}\right)  \left( \frac{1}{{K_u}}-\frac{1}{{K_s}}\right) }{\frac{1}{{K_d}}-\frac{1}{{K_u}}}
\end{split}
\label{eq:el:conc:M}
\end{equation}

\hl{In addition, the properties $U_i$ (Eq. (\ref{eq:el:UiDi})), the undrained elastic stiffness matrix $M^u_{ij}$ (Eq. (\ref{eq:el:Mu})), its inverse $C^u_{ij}$, and the parameters $E_{u,i}$ and $\nu_{u,i}$ (Eq. (\ref{eq:el:Eu_nuu})) can be determined.}
\hl{We obtain hence a complete set of parameters depending on the \hz{nine} unknowns ${E_z^0}$, $E_z^\infty $, $\beta$, $R_E$, $\nu_{hz}$, $\nu_{hh}$, $R_H$, $K_s$ and $K_u$.
Here we assume a priori that the coefficients $R_E$, $R_H$, $\nu_{hz}$, $\nu_{hh}$, $K_s$ and $K_u$ are constant with $\sigma'$.}
\hl{Based on these \hz{nine} coefficients, we can evaluate the theoretical values of the 14 parameters which were investigated in the experiments ($D_i$, $H_i$, $U_i$, $E_i$, $\nu_{i}$, $E_{u,i}$ and $\nu_{u,i}$) and compare them to the measured data.}

\hl{For a given set of \hz{nine} unknown coefficients, we can compute the sum of the squared relative errors between all our measured data and the calculated dataset. Due to the fact that different parameters with different units were fitted here, we used the method of squared relative errors. To minimize the sum of residuals we used a Python SciPy differential evolution algorithm \mbox{\citep{Storn1997,2020SciPy-NMeth}}, which found a global minimum with the values for the \hz{nine} unknowns presented in Tab. \ref{tab:el:summary_results_practical}. The result of the fitting is displayed together with the experimental data in the respective Figures \ref{fig:el:Kdsig}, \ref{fig:el:Hsig} and \ref{fig:el:Ku} - \ref{fig:el:vu}. For each curve obtained by the regression, we show its standard deviation $S$ in the Figures. }  

\begin{table*}
\caption{\hl{Summary of the best-fit parameter set. The independent variables were varied in the fitting procedure. We also present values of other resulting parameters with their stress dependency, which are useful for common numerical applications.}}
\label{tab:el:summary_results_practical}       % Give a unique label
\renewcommand\arraystretch{1.2} 
\fcolorbox{white}{white}{%
\begin{tabular}{lllll}
\hline\noalign{\smallskip}
 &						&Fit	&\multicolumn{2}{l}{Resulting parameters in function of $\sigma'$ [MPa]}        \\ 
\noalign{\smallskip}\hline\noalign{\smallskip}
${E_z^0}$&[GPa]			&0.30	&&\\
$E_z^\infty $&[GPa]		&3.68	&&\\
$\beta$&[MPa$^{-1}$]	&0.34	&&\\
$E_z$ &[GPa]	    	&		& $ \left[  3.06 \, \exp (-0.34 \, \sigma') +0.27 \right] ^{-1} $			&[Eq.(\ref{eq:el:Kdmodel})] \\
$R_E$ &[-]				&1.84	&&\\
$E_h$ &[GPa]			&		&	$1.84 \, \left[  3.06 \, \exp (-0.34 \, \sigma') +0.27 \right] ^{-1} $		&[Eq.(\ref{eq:el:R_E})]	\\
$\nu_{zh}$ &[-] 	    &0.11	&&\\                                            
$\nu_{hh}$ &[-]       	&0.29	&&\\  
$R_H$ &[-]				&2.95	&&\\  
$K_s$ &[GPa]			&19.69	&&\\                                          
$b_h$ &[-]          	&		& $ \left[ \exp (0.34 \, \sigma')  +13.02 \right]  \cdot  \left[ 1.18 \, \exp (0.34 \, \sigma')  +13.22 \right] ^{-1}$ &[Eq.(\ref{eq:el:bh})]	\\
$b_z$ &[-]          	&		& $\left[ \exp (0.34 \, \sigma')+13.07 \right]  \cdot  \left[  1.14 \, \exp (0.34 \, \sigma')+12.93  \right] ^{-1}$ &[Eq.(\ref{eq:el:bz})]	\\
$K_u$ &[GPa]       		&10.43	&&\\ 
$M$ &[GPa]          	&		&$\left[    10.38 \, \exp ({1.02 \, \sigma' })+2.77 \cdot{{10}^{2}}\, \exp ({0.68 \,\sigma' })+1.79 \cdot{{10}^{3}}\, \exp ({0.34 \,\sigma' })   \right]   $			&[Eq.(\ref{eq:el:conc:M})] \\
&					&&$\times \left[    \, \exp ({1.02 \, \sigma' })+0.26 \cdot{{10}^{2}}\, \exp ({0.68 \,\sigma' })+1.71 \cdot{{10}^{2}}\, \exp ({0.34 \,\sigma' })-1.89  \right] ^{-1}  $&\\
\noalign{\smallskip}\hline                                     
\end{tabular}%
}
\end{table*}

\hl{The data on $K_d$ and $H$ (Fig. \ref{fig:el:Kdsig}, \ref{fig:el:Hsig}) show a good fit, with low values of standard deviation (0.3 and 0.4 GPa, respectively). Here the fit gives a high confidence due to the largest number of experimental data. }
\hl{Moreover, the a priori assumption of constant $K_s$ appeared sufficient for reproducing the relationship between the two parameters $K_d$ and $H$ (Eq. (\ref{eq:el:2})). The obtained value of $K_s=$ 19.69 GPa is very close to the data provided by \mbox{\cite{Belmokhtar201787}} of $K_s \approx $ 21 GPa.} 
\hl{In the case of $K_u$, we adopted a constant parameter, shown in Fig. \ref{fig:el:Ku}. Note that due to some significant outliers in the experimental data, the fitted value $K_u=$ 10.43 GPa shows a high standard deviation of $S=$ 3.4 GPa.} \hz{Moreover, the fact that the experimental data are not symmetrically distributed around the fitted function indicates a fairly poor fit.}

\hl{The standard deviation $S$ of the regression for $D_h$, $D_z$, $H_h$ and $H_z$ (Fig. \ref{fig:el:Daniso}, \ref{fig:el:Haniso}) was found equal to 2.0, 0.5, 2.5 and 0.6 GPa, respectively. These values indicate acceptable fitting. For the parameters in $h$ direction, $S$ is generally higher due to the higher absolute values.}
\hl{For $U_h$ and $U_z$, the values of $S$ were notably higher ($S=$ 15.5 GPa and 6.7 GPa, respectively), due to the high dispersion of the undrained measurements (Fig. \ref{fig:el:Kuaniso}).}

\hl{One can see that the fitted relationships for the deviatoric parameters $E_{i}$ and $\nu_{i}$ (Fig. \ref{fig:el:Edrained}, \ref{fig:el:vdrained}) simulate qualitatively the measured anisotropic \hz{and stress dependent} characteristics of the COx claystone. However, quantitatively we observe a \hz{significant} underestimation \hz{of the measured data}, resulting in \hz{very large} standard deviations $S$ of $E_{z}$ and $\nu_{hz}$ with values of 3.0 GPa and 0.04, respectively.}
\hz{These large standard deviations are also due to the relatively small number of measurements under triaxial stress conditions. For $E_{h}$ and $\nu_{hh}$, the number of measurements is not sufficient to calculate the standard deviation of the estimate.}

\hl{Also the obtained relationship for $E_{u,z}$ shows a large standard deviation $S=$ 1.5 GPa (Fig. \ref{fig:el:Eu}). In parallel bedding direction, we measured a much larger $E_{u,h}$ which is \hz{cannot} be reproduced by the regression. \hz{In this direction, the number of measurements is not sufficient for calculating the standard deviation, but we can observe a poor fit in the Figure.}} 
\hl{In the experiments we measured relatively high undrained Poisson ratios, larger than 0.35 (Fig. \ref{fig:el:vu}). Remarkably, the results of the fitted parameter set show similar values, indicating that these high values are compatible with the poroelastic framework. However, the standard deviation of these regressions is rather high ($S=$ 0.14 for $\nu_{u,hz}$, \hz{no quantitative measure for $\nu_{u,hh}$ due to lack of data-points}).}

\section{Discussion}
\label{sec:el:discuss}

\hz{The proposed multivariate regression scheme is a useful tool to fit multiple dependent variables (experimental material parameters) through multiple independent variables (model parameters) and to analyse a complex set of transversely isotropic poroelastic parameters. For properties measured under isotropic loading conditions, the model shows a good fit due to the large number of experimental data.} 
Some compatibility issues were encountered in the response to deviatoric loads, where the model \hz{significantly} underestimates the measured moduli $E_{i}$ and $E_{u,i}$. 
This could be due to the fact that we express the moduli \hz{only} \hl{as a} function of effective \hl{isotropic} stress, which is not sufficient for reproducing the observed behaviour \hz{in deviatoric tests.} 
\hz{Some studies (e.g. \mbox{\citealp{Cariou201236,Zhang201912}}) reported an increase of the Young modulus as a function of $q$ during tests under partially saturated conditions. Additional experimental results and the consideration of a parameter dependency on $q$ could improve the set of material coefficients.}

\begin{figure}[htbp]
  \includegraphics[width=\columnwidth]{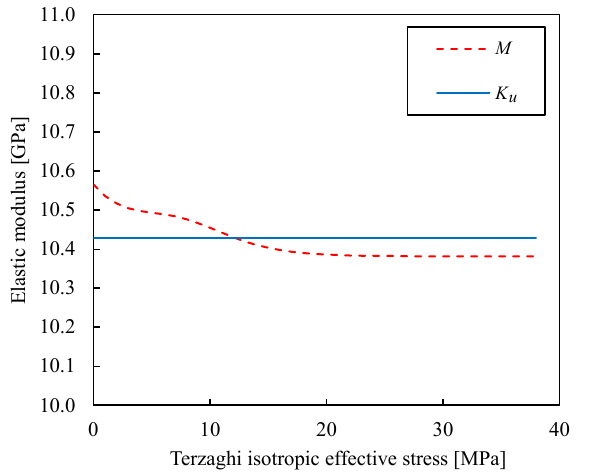}
\caption{\hl{Biot's modulus $M$ compared with the undrained bulk modulus $K_u$, both evaluated through the regression analysis.}}
\label{fig:el:M}       % Give a unique label
\end{figure}

\hl{We are also able to evaluate several other poroelastic characteristics, not measured experimentally, by using the fitted dataset. For instance, we can calculate the Biot modulus $M$ through Eq. (\ref{eq:el:conc:M}), presented in Fig. \ref{fig:el:M} and Tab. \ref{tab:el:summary_results_practical}. This coefficient, often used as a model parameter in engineering applications, changes very little with effective stress and is close to the undrained bulk modulus $K_u$.} 

\begin{figure}[htbp]
  \includegraphics[width=\columnwidth]{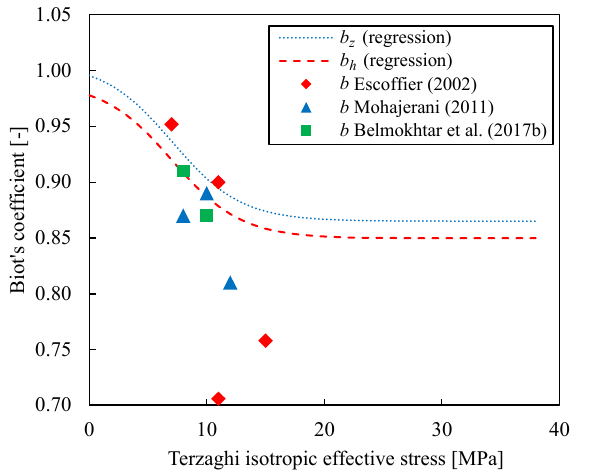}
\caption{\hl{Biot's effective stress coefficients $b_i$ obtained through the regression, compared with literature data on the isotropic parameter $b$. Two datapoints from \mbox{\cite{Escoffier2002}} for $b= 0.54$ and 0.32, both at 23 MPa effective stress, are not shown in the figure.}}
\label{fig:el:b}       % Give a unique label
\end{figure}

\hl{The Biot coefficients $b_h$ and $b_z$ were calculated using Eq. (\ref{eq:el:bh}) and (\ref{eq:el:bz}). One can compute stress dependent Biot's coefficients, presented in Fig. \ref{fig:el:b} and Tab. \ref{tab:el:summary_results_practical}. Both coefficients show values close to 1.0 for low effective stress, decrease slightly with effective stress and remain constant above around 14 MPa effective stress. At the in-situ effective stress of around 10 MPa, we obtain values $b_i$ close to 0.9, which is agreement with previous studies (Fig. \ref{fig:el:b}). Interestingly, a less pronounced anisotropy can be seen here, compared to other elastic coefficients. High values of $b_i$ originate from small differences between $K_d$ and $H$, while a small anisotropy is due to small differences between $D_h$ and $H_h$, and between $D_z$ and $H_z$.} 

\begin{figure}[htbp]
  \includegraphics[width=\columnwidth]{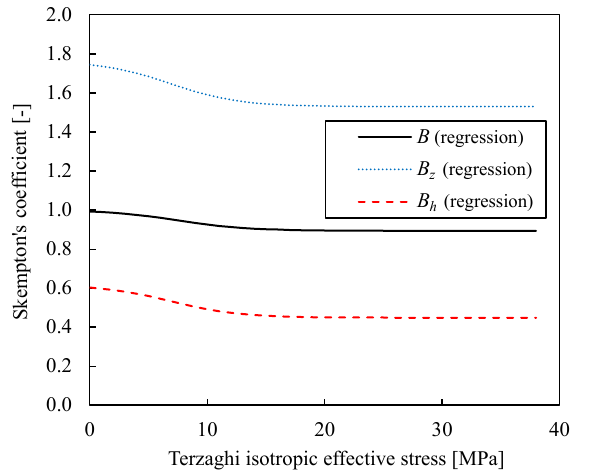}
\caption{Skempton's coefficients $B$ and $B_i$ evaluated through our regression analysis.}
\label{fig:el:B}       % Give a unique label
\end{figure}

\hl{The isotropic Skempton coefficient $B$ (Fig. \ref{fig:el:B}), calculated through Eq. (\ref{eq:el:B}), shows relatively high values between 0.9 and 1.0.  Similar to $b_i$, this coefficient decreases slightly with effective stress up to 14 MPa. Also \mbox{\cite{Belmokhtar201787}} and \mbox{\cite{Mohajerani201211,Mohajerani201413}} determined rather high coefficients $B$ with values of 0.87 and 0.84, respectively.
Using Eq. (\ref{eq:el:Bi}), one is able to calculate the anisotropic Skempton's coefficients, presented in Fig. \ref{fig:el:B}. 
Note that here, as well as for $b_i$, a combination of different involved regressions increases the uncertainty of the calculated relationships. One observes a large difference between $B_z$ and $B_h$, with values of 1.6 and 0.5, respectively, at 10 MPa Terzaghi effective stress. Such significant anisotropies of the Skempton coefficient have been addressed by \mbox{\cite{Holt2018}}, who tested soft shales with $B$ close to 1.0. They found properties equivalent to $B_z$ around 1.5 and $B_h$ around 0.6, similar to the findings of this study. These authors were able to back-calculate their measurements by adopting a transversely isotropic poroelastic model. Moreover, they emphasized the significance of anisotropic $B_i$ on pore pressure increase due to subsurface drilling, injection or depletion. In the numerical study of \mbox{\cite{Guayacan-Carrillo201701}}, transversely isotropic material parameters provided the best results to model the in-situ measurements of induced pore pressures during excavation of galleries in the COx claystone. Although not discussed explicitly, the material properties used in their study most likely resulted in anisotropic Skempton's parameters, able to reproduce the non-uniform pore pressure field observed around the galleries.}

\section{Conclusion}
\label{sec:el:6}
A laboratory study using experimental equipment and testing procedures adopted for very low permeability materials allowed us to conduct a \hl{series of time efficient} isotropic and deviatoric experiments on the COx claystone. Precise strain gage measurements were employed to investigate its transversely isotropic poromechanical properties and to establish a set of independent poroelastic parameters.  

\begin{itemize}
  \item Tests on different \hl{specimens} from three cores provided consistent poromechanical coefficients, which confirms the reproducibility of the experiments, the same \hl{specimen} qualities and shows no noticeable natural variability between the cores.
  
\item \hl{The measurements show a clear stress dependency of the drained bulk modulus $K_d$ and the Biot modulus $H$, the drained and undrained Young moduli and a less pronounced stress dependency of the undrained bulk modulus and the undrained Young modulus. The drained and undrained Poisson ratios perpendicular to the isotropy plane were observed to change very little with effective stress.} 

\item The decrease of the drained bulk modulus with decreasing effective stress was observed to be reversible in isotropic tests. The reduction of moduli could probably be attributed to an elastic opening of microcracks.

\item \hl{A remarkable anisotropy was found for several elastic coefficients. The drained Young moduli showed a anisotropy ratio close to 1.8, while the isotropic stress parameters $D_i$ and $H_i$ had a ratio close to 3.0 (ratio between modulus parallel and perpendicular to bedding). Drained Poisson's ratios were close to 0.3 parallel to bedding, whereas perpendicular to bedding we found values around 0.15. In undrained conditions, the anisotropy of moduli was less developed. Relatively high undrained Poisson's ratios around 0.4 illustrate peculiar undrained characteristics.}

\item We carried out \hl{a regression analysis}, which aimed to match a stress dependent set of transversely isotropic poroelastic coefficients to the laboratory measurements. \hl{The obtained \hz{seven} independent coefficients (9 coefficients due to stress dependency}, Tab. \ref{tab:el:summary_results_practical}) could \hl{satisfactorily} represent the measured parameters \hz{under isotropic loading conditions}. Some compatibility issues were encountered, where elastic properties evaluated from deviatoric tests tend to be stiffer than those back-calculated from the fitted parameter set. 
\item \hz{The observed poor fit for parameters measured under deviatoric loading conditions could be improved by additional experimental data, by taking into account 
a parameter dependency on deviatoric stress or considering other effects, which are not captured by our transversely isotropic poroelasticity assumption with parameters depending on isotropic stress only. }

\item \hl{The fitted unjacketed compression modulus, which is assumed to be isotropic at the micro-scale and constant with effective confining stress, gave a value of $K_s=19.69$ GPa. This is in accordance with the direct measurements of \mbox{\cite{Belmokhtar201787}}.} 

\item \hl{A negligible anisotropy of the calculated Biot's coefficients $b_h$ and $b_z$ was indicated by the fitted parameters. Relatively high values of the Biot coefficients highlight the importance of hydromechanical couplings in this material. 
The calculated Skempton's coefficients showed a significant anisotropy, with higher values perpendicular to bedding.} 
\end{itemize}

The precise knowledge of the poromechanical properties of the COx claystone is important not only for the hydromechanical modelling of the rock behaviour around excavated drifts \citep{Guayacan-Carrillo201701}, but also for the analysis of the deformations due to thermally induced pore pressures \citep{Gens200720,Garitte201701}. The parameters evaluated in this study give confidence due to their \hl{inter-}compatibility, evidencing a clear
anisotropy. 
The observed stress dependency of elastic properties could be considered in in-situ modelling. The material was seen to become more compliant under decreasing effective stresses, which increases the elastic deformations in unloading paths.
%As quoted by \cite{Gens200720} and \cite{Garitte201701}, a smaller stiffness reduces the induced pore pressure during thermal pressurization, as the solid becomes more compliant to the increasing volume of pore fluid. However, at the same time the thermally induced strains in undrained conditions will increase. 

\section*{Conflict of interest}

The authors declare that there are no known conflicts of interest associated with this publication.

\bibliographystyle{spbasic}      % basic style, author-year citations
\bibliography{bib_article}   % name your BibTeX data base

%\begin{thebibliography}{}
%\end{thebibliography}

\end{document}